\documentclass[aps,11pt,nofootinbib,prd,longbibliography]{revtex4-1}

\usepackage{feynmp-auto}
\usepackage{amsmath}
\usepackage[utf8x]{inputenc} 

\usepackage{bbm}
\usepackage{soul}	
\usepackage{xcolor}

\usepackage{epsfig}
\usepackage{color}
\usepackage{slashed}
\usepackage{comment}
\usepackage{epstopdf}

\usepackage{array}
\usepackage{booktabs}
\usepackage{mathrsfs}
\usepackage[euler]{textgreek}
\usepackage{amsmath}
\usepackage{amsthm}
\usepackage{amsfonts}
\usepackage{amssymb}
\usepackage{graphicx}
\graphicspath{ {Figures/} }
\usepackage[font={small}]{caption}
\usepackage{float}
\usepackage{multirow}
\usepackage[export]{adjustbox}
\usepackage[hang, flushmargin,bottom]{footmisc} 
\usepackage{hyperref}
\hypersetup{
     colorlinks   = true,
     citecolor    = blue
}

\usepackage{placeins}
\usepackage{enumitem}

\usepackage{slashed}
\usepackage{bbm}
\usepackage{appendix}

\usepackage{titlesec}
\titleformat{\chapter}[display]
  {\normalfont\LARGE\bfseries}
  {\chaptertitlename\ \thechapter}{5pt}{\LARGE}
  \titlespacing*{\chapter}{0pt}{-20pt}{35pt}
\usepackage{bigstrut}
\usepackage{pst-node}
\usepackage{pstricks}

\usepackage{physics}
\usepackage{mathtools}
\usepackage{setspace}
\usepackage{fancyhdr}
\usepackage[makeroom]{cancel}

\newcommand{\be}{\begin{equation}}
\newcommand{\ee}{\end{equation}}
\newcommand{\bes}{\begin{equation*}}
\newcommand{\ees}{\end{equation*}}

\usepackage{hyperref}
\hypersetup{%
  colorlinks = true,
  linkcolor  = black
}
\usepackage{soul}

\newcommand{\beq}{\begin{equation}}
\newcommand{\eeq}{\end{equation}}

\newcommand{\SU}{\,{\rm SU}}
\newcommand{\U}{\,{\rm U}}

\usepackage[normalem]{ulem}

\DeclareUnicodeCharacter{2212}{-}

\AtBeginDocument{%
    \newwrite\bibnotes
    \def\bibnotesext{Notes.bib}
    \immediate\openout\bibnotes=\jobname\bibnotesext
    \immediate\write\bibnotes{@CONTROL{REVTEX41Control}}
    \immediate\write\bibnotes{@CONTROL{%
    apsrev41Control,author="08",editor="1",pages="1",title="0",year="1"}}
     \if@filesw
     \immediate\write\@auxout{\string\citation{apsrev41Control}}%
    \fi
}%

\usepackage{amsfonts}
\usepackage{graphicx}
\usepackage{tcolorbox}
\usepackage{marvosym} 
\usepackage{tikz}
\usepackage{tabulary}
\usepackage{pgfplots}
\usepackage{subfig}
\usepackage{amsmath}
\usepackage{slashed}
\usepackage{mathtools}
\usepackage{nccmath}
\usepackage{verbatim}
\usepackage{color}
\usepackage{amsmath, amsthm, amssymb, amsfonts}
\usepackage{array}
\newcolumntype{P}[1]{>{\centering\arraybackslash}p{#1}}
\usepackage{enumerate}
\usepackage{physics}
\usepackage[font={small, up}]{caption}
\usepackage{tikz}
\usetikzlibrary{"arrows", "automata", "backgrounds", "calendar", "chains", "matrix", "mindmap", "patterns", "petri", "shadows", "shapes.geometric", "shapes.misc", "spy", "trees"}
\graphicspath{{./images/}}
\usepackage{appendix}

\NewDocumentCommand\semiloop{O{black}mmmO{}O{above}}
{%
\draw[#1] let \p1 = ($(#3)-(#2)$) in (#3) arc (#4:({#4+180}):({0.5*veclen(\x1,\y1)})node[midway, #6] {#5};)
}

\NewDocumentCommand\hello{O{black}mmmO{}O{above}}
{%
\draw[#1] let \p1 = ($(#3)-(#2)$) in (#3) arc (#4:({#4-180}):({0.5*veclen(\x1,\y1)})node[midway, #6] {#5};)
}

\tikzstyle{block} = [draw, rectangle, 
    minimum height=3em, minimum width=6em]

\usetikzlibrary{fit}
\tikzset{%
  highlight/.style={rectangle,rounded corners,fill=red!15,draw,
    fill opacity=0.5,thick,inner sep=0pt}
}
\tikzset{%
  highlightblue/.style={rectangle,rounded corners,fill=blue!15,draw,
    fill opacity=0.5,thick,inner sep=0pt}
}

\newcommand\hlight[1]{\tikz[overlay, remember picture,baseline=-\the\dimexpr\fontdimen8\textfont2\relax]\node[rectangle,fill=gray!50,rounded corners,fill opacity = 1,draw,thick,text opacity =1] {$#1$};}

\usepackage{hyperref}	

\begin{document}
\title{\Large {\bf{Leptoquarks and Matter Unification: \\ Flavor Anomalies and the Muon $g-2$}}}
\author{Pavel Fileviez P\'erez$^{1}$, Clara Murgui$^{2}$, Alexis D. Plascencia$^{1}$}
\affiliation{$^{1}$Physics Department and Center for Education and Research in Cosmology and Astrophysics (CERCA), 
Case Western Reserve University, Cleveland, OH 44106, USA \\
$^{2}$Walter Burke Institute for Theoretical Physics, California Institute of Technology, Pasadena, CA 91125}
\email{pxf112@case.edu, cmurgui@caltech.edu, alexis.plascencia@case.edu}
\vspace{1.5cm}

\begin{abstract}
We discuss the minimal theory for quark-lepton unification at the low scale. In this context, the quarks and leptons are unified in the same representations 
and neutrino masses are generated through the inverse seesaw mechanism. The properties of the leptoquarks predicted in this theory are discussed in detail 
and we investigate the predictions for the leptonic and semi-leptonic decays of mesons. We study the possibility to explain the current value of 
$\mathcal{R}_K$ reported by the LHCb collaboration and the value of the muon anomalous magnetic moment reported by the Muon $g-2$ experiment at Fermilab.
\end{abstract}

\maketitle
\hypersetup{linkcolor=blue}
\section{INTRODUCTION}
The idea of quark-lepton unification by J. Pati and A. Salam~\cite{Pati:1974yy} provides a simple and elegant approach to think about unification of matter and interactions in nature. The minimal theory of Pati-Salam is very predictive because it predicts, at the scale where matter unifies, that the masses of the charged leptons and down-quarks are equal, and the masses for the up-quarks are equal to the Dirac masses for neutrinos. Moreover, the $\SU(4)_C$ gauge symmetry must be broken around the canonical seesaw scale, $10^{14}$ GeV, in order to achieve small neutrino masses, via the seesaw mechanism~\cite{Minkowski:1977sc,Yanagida:1979as,GellMann:1980vs,Mohapatra:1979ia}, in agreement with experiments. 

The simplest quark-lepton unification theory that can be realized at the TeV scale was proposed in Ref.~\cite{Perez:2013osa}. This theory is based on the $\SU(4)_C \otimes \SU(2)_L \otimes \U(1)_R$ gauge group and in order to have a consistent theory for fermion masses at the low 
scale, neutrino masses are generated through the inverse seesaw mechanism~\cite{Mohapatra:1986aw,Mohapatra:1986bd}. 
 This theory for quark-lepton unification predicts, among the new fields required for its consistency, the existence of mediators that interact simultaneously with both leptons and quarks. Particularly, this theory predicts a vector leptoquark, $X_\mu \sim (\mathbf{3},\mathbf{1},2/3)_{\rm SM}$, and two scalar leptoquarks, $\Phi_3 \sim (\mathbf{\bar{3}},\mathbf{2},-1/6)_{\rm SM}$ and $\Phi_4\sim (\mathbf{3},\mathbf{2},7/6)_{\rm SM}$, which mediate exotic processes that could otherwise not be seen in the context of the Standard Model (SM)\footnote{We label with the subscript {\rm SM} the quantum numbers of the corresponding field under the Standard Model gauge group $\SU(3)_C \otimes \SU(2)_L \otimes \U(1)_Y$.}. For a review about the phenomenology of leptoquarks see Ref.~\cite{Dorsner:2016wpm}.

Recently, the LHCb collaboration has reported results for the ratio $\mathcal{R}_K$ defined as
\begin{equation*}
    {\cal R}_{K} =  \frac{{\rm Br}(B^+ \to K^+ \mu^+ \mu^-)}{{\rm Br}(B^+ \to K^+ e^+ e^-)},
\end{equation*}
which is predicted to be 1 in the SM. The measurement reported by LHCb~\cite{Aaij:2021vac} using data from Run 2 is
\begin{equation}
{\cal R}_K^\text{exp}(1.1 < q^2 < 6.0 \text{ GeV}^2)=0.846^{+0.042 \, + 0.013}_{-0.039 \, -0.012},
\label{eq:RKexp}
\end{equation}
which is in tension with the SM prediction at $3.1\sigma$. This observable, together with ${\cal R}_{K^*}$ and the leptonic decays $\text{Br}(B_s \to \ell^+ \ell^-)$, where $\ell=e,\mu$, are usually classified as {\it clean observables}; in the former the hadronic and the long distance effects cancel almost exactly. These experimental results have motivated many studies in the particle physics community, see e.g.~\cite{Gripaios:2014tna,Alonso:2015sja,Calibbi:2015kma,Pas:2015hca,Becirevic:2016oho,Barbieri:2016las,Crivellin:2017zlb,Capdevila:2017bsm,DAmico:2017mtc,Hiller:2017bzc,Buttazzo:2017ixm,Assad:2017iib,DiLuzio:2017vat,Calibbi:2017qbu,Faber:2018qon,Balaji:2018zna,Fornal:2018dqn,Popov:2019tyc,Balaji:2019kwe,Angelescu:2021lln,Hiller:2021pul,Altmannshofer:2021qrr,Cornella:2021sby,Fleischer:2021yjo,Lancierini:2021sdf}. 

There are also new results reported by the Fermilab Muon $g-2$ experiment on the anomalous magnetic moment of the muon $a_\mu$ from their Run 1~\cite{Abi:2021gix}. The combined result with the one from the E821 experiment at BNL~\cite{Bennett:2006fi} deviates from the SM prediction\footnote{There are results from lattice QCD that are compatible with the experimental value~\cite{Borsanyi:2020mff}.} by $4.2\sigma$
\begin{equation}
\label{eq:FermiLab}
\Delta a_\mu = a_\mu^{\rm exp} - a_\mu^{\rm SM} = (251 \pm 59 ) \times 10^{-11}.
\end{equation}
There have been different proposals of theories beyond the SM to explain this anomaly, see e.g. Refs.~\cite{Freitas:2014pua, Chiu:2014oma, Calibbi:2018rzv, Kowalska:2018ulj, Mandal:2019gff, Dorsner:2019itg,Bigaran:2020jil, Crivellin:2020tsz, Capdevilla:2021rwo, Baker:2021yli, Chiang:2021pma, Zhu:2021vlz, Buen-Abad:2021fwq, Amaral:2021rzw, Bai:2021bau, Athron:2021iuf}. These results for $\mathcal{R}_K$ and $(g-2)_\mu$, which are naturally expected if new physics is around the multi-TeV scale, could help us find a new direction for physics beyond the Standard Model. 

In this article, we investigate the possibility to explain the experimental value of $\mathcal{R}_K$ in two main scenarios. In the first scenario the scalar leptoquark $\Phi_3 \sim (\mathbf{\bar{3}},\mathbf{2},-1/6)_{\rm SM}$ gives the main contribution to the relevant meson decays, while in the second scenario the scalar leptoquark $\Phi_4\sim (\mathbf{3},\mathbf{2},7/6)_{\rm SM}$ plays the main role to explain the value of $\mathcal{R}_K$. In the second scenario, $\Phi_4$ couples mostly to electrons as required by constraints from lepton flavor violation. We also show that the component $\phi_4^{5/3}$ of $\Phi_4$, due to the enhancement by the factor $m_t/m_\mu$, can explain the reported value of the anomalous magnetic moment of the muon by the Muon $g-2$ collaboration. 
Moreover, we show that the theory of minimal quark and lepton unification can address simultaneously the anomaly in $\mathcal{R}_{K^{(*)}}$ and the experimental value of $(g-2)_\mu$. In these scenarios, $\Phi_3$ and $\Phi_4$ explain the flavor anomalies while $\Phi_4$ addresses the $(g-2)_\mu$ through its couplings mainly to muons, so that the predictions are consistent with constraints from lepton flavor violation.

This article is organized as follows: in Section~\ref{sec:theory} we discuss the minimal theory for quark-lepton unification at the low scale, in Section~\ref{sec:Bdecays} we investigate the predictions for meson decays and discuss the different leptoquark candidates to explain the experimental value of $\mathcal{R}_K$. In Section~\ref{sec:muong2} we discuss the possibility to explain the recent experimental results for the $g-2$ of the muon and demonstrate that the theory can address the anomalies in the clean observables involving $b\to s$ transitions and the muon $g-2$. Finally, in Section~\ref{sec:summary} we summarize our main findings.  

\section{MINIMAL THEORY FOR QUARK-LEPTON UNIFICATION}
\label{sec:theory}
The minimal theory for quark-lepton unification that can describe physics at the TeV scale was proposed in Ref.~\cite{Perez:2013osa}.
This theory is based on the gauge symmetry, $${\cal G}_{QL}=\SU(4)_C \otimes \SU(2)_L \otimes \U(1)_R,$$ and the SM matter fields are unified as
\begin{align}
F_{QL} =
\left(
\begin{array}{cc}
u 
&
\nu 
\\
d 
&
e
\end{array}
\right) \sim (\mathbf{4}, \mathbf{2}, 0),
\, \, & \hspace{1cm}
F_u =
\left(
\begin{array}{cc}
u^c 
&
\nu^c
\end{array}
\right) \sim (\mathbf{\bar{4}}, \mathbf{1}, -1/2), \\[0.5ex]
\, \, {\rm{and}} \, \,  &
\hspace{1cm} F_d =
\left(
\begin{array}{cc}
d^c 
&
e^c
\end{array}
\right) \sim (\mathbf{\bar{4}}, \mathbf{1}, 1/2).
\end{align}
Here all the SM fields and $\nu^c$ are in the left-handed representation, and the unification for quarks and leptons is for each SM family. This theory can be seen as a low energy limit of the Pati-Salam model based on $\SU(4)_C \otimes \SU(2)_L \otimes \SU(2)_R$, when the $\SU(2)_R$ symmetry is broken to $\U(1)_R$. One can also obtain the gauge symmetry ${\cal G}_{QL}$ from a unified theory based on $\SU(6)$.

In this theory the gauge fields live in 
\begin{eqnarray}
A_\mu =
\left(
\begin{array} {cc}
G_\mu & X_\mu/\sqrt{2}  \\
X_\mu^*/\sqrt{2} & 0  \\
\end{array}
\right) + T_4 \ B_\mu^{'} \sim (\mathbf{15}, \mathbf{1},0),
\end{eqnarray}
where $G_\mu \sim (\mathbf{8},\mathbf{1},0)_\text{SM}$ are the gluons, $X_\mu \sim (\mathbf{3},\mathbf{1},2/3)_\text{SM}$ are vector leptoquarks, and $B_\mu^{'} \sim (\mathbf{1},\mathbf{1},0)_\text{SM}$. The Higgs sector is composed of three scalar representations: 
\begin{eqnarray}
H_1^T & = & \left( H^+   \  H^0 \right) \sim (\mathbf{1}, \mathbf{2}, 1/2), \, \,\,\,\,\,
\chi = \left(  \chi_u  \  \ \chi_R^0 \right) \sim (\mathbf{4}, \mathbf{1}, 1/2), \, \,\,\,\,\, {\rm{and}} \nonumber \\[1ex]
\Phi &=& 
\left(
\begin{array} {cc}
\Phi_8 & \Phi_3  \\
\Phi_4 & 0  \\
\end{array}
\right) + \sqrt{2} \, T_4 \ H_2 \sim (\mathbf{15}, \mathbf{2}, 1/2).
\end{eqnarray}
Here $H_2 \sim (\mathbf{1}, \mathbf{2}, 1/2)_\text{SM}$ is a second Higgs doublet, $\Phi_8 \sim (\mathbf{8}, \mathbf{2}, 1/2)_\text{SM}$, and the scalar leptoquarks  
$\Phi_3 \sim (\mathbf{\bar{3}}, \mathbf{2},-1/6)_\text{SM}$ and  $\Phi_4 \sim (\mathbf{3}, \mathbf{2}, 7/6)_{\rm SM}$. The $T_4$ generator of $\SU(4)_C$ in the above equation is normalized as
$
T_4 =
\frac{1}{2 \sqrt{6}} \rm{diag} (1,1,1,-3).
$
The ${\cal G}_{QL}$ gauge group is spontaneously broken to the SM gauge group by the vacuum expectation value (VEV) of the scalar
field $\langle \chi_R^0 \rangle =v_{\chi}/\sqrt{2}$, which gives mass to the vector leptoquark $X_\mu$, defining the scale of matter unification.

The Yukawa interactions in this theory are given by
\begin{eqnarray}
- {\cal L}_{QL}^{Y} &=&
Y_1  \, {F}_{QL} F_u H_1  \ + \ Y_2 \,  {F}_{QL} F_u \Phi  \
+ \  Y_3 \,  H_1^\dagger {F}_{QL} F_d  \ + \  Y_4  \, \Phi^\dagger {F}_{QL}  F_d   + \mbox{h.c.},
\label{eq:Yukawa}
\end{eqnarray}
and the mass matrices for the SM fermions read as
\begin{eqnarray}
M_U &=& Y_1 \frac{ v_1}{\sqrt{2}} + \frac{1}{2 \sqrt{3}} Y_2 \frac{ v_2}{\sqrt{2}}, \quad \quad 
M_\nu^D = Y_1 \frac{ v_1}{\sqrt{2}} - \frac{\sqrt{3}}{2 } Y_2 \frac{ v_2}{\sqrt{2}}, \label{eq:fermionmasses1}\\
M_D &=& Y_3  \frac{v_1}{\sqrt{2}} + \frac{1}{2 \sqrt{3}} Y_4 \frac{ v_2}{\sqrt{2}}, \quad \quad
\, \, M_E = Y_3\frac{ v_1}{\sqrt{2}} - \frac{\sqrt{3}}{2} Y_4 \frac{ v_2}{\sqrt{2}}.
\label{eq:fermionmasses2}
\end{eqnarray}
Here the VEVs of the Higgs doublets are defined as $\langle H^0_1 \rangle = v_1 / \sqrt{2}$, and $\langle H^0_2  \rangle  = v_2/\sqrt{2}$.
Notice that without the scalar field $\Phi$ one cannot generate a consistent relation for charged fermion masses.
Now, in order to generate small neutrino masses at the low scale one needs to go beyond the canonical seesaw mechanism.
We can generate small Majorana  masses for the light neutrinos if we add three new singlet left-handed fermionic fields $S\sim (\mathbf{1}, \mathbf{1},0)$ and use the following interaction terms~\cite{Perez:2013osa}, which emerge in the Lagrangian once the fermion singlets are included,
\begin{eqnarray}
- {\cal L}_{QL}^\nu &=&
Y_5 F_u \chi S  \ + \  \frac{1}{2} \mu S S   + \mbox{h.c.}.
\end{eqnarray} 
In this case the mass matrix for neutrinos in the basis ($\nu$, $\nu^c$, $S$) reads as
\begin{equation}
\left( \nu \  \nu^c \  S  \right) 
\left(\begin{array}{ccc} 
0 & M_\nu^D & 0  \\ 
(M_\nu^D)^T & 0 & M_\chi^D \\
0 &  (M_\chi^D)^T & \mu
\end{array}\right)  
\left(\begin{array}{c} \nu \\  \nu^c \\ S  \end{array}\right).
\end{equation}
Here $M_\nu^D$ is given by Eq.~\eqref{eq:fermionmasses1} and 
$
M_\chi^D = Y_5 \, v_\chi / \sqrt{2}.
$
The light neutrino mass is given by 
\begin{equation}
m_\nu \approx \mu \, (M_\nu^D)^2 / (M_\chi^D)^2,
\end{equation}
if $M_\chi^D \gg M_\nu^D \gg \mu$ holds. Such hierarchy is motivated by the different scales of the theory: $M_\chi^D \propto v_\chi$, which determines the scale of matter unification, $M_\nu^D \propto v_{1,2}$, which defines the electroweak scale, and $\mu$ is instead protected by a fermion symmetry, so that it is technically natural to assume it small.
Notice that neutrinos would be massless in the limit $\mu \rightarrow 0$, which is the usual relation in the inverse seesaw mechanism.

The vector leptoquarks, $X_\mu \sim (\mathbf{3}, \mathbf{1}, 2/3)_{\rm SM}$, have the following interactions
\begin{equation}
{\cal{L}} \supset \frac{g_4}{\sqrt{2}} X_\mu \left(  \bar{Q}_L  \gamma^\mu \ell_L +  \bar{u}_R \gamma^\mu \nu_R +  \bar{d}_R \gamma^\mu e_R  \right) 
 +  \rm{h.c.},
\end{equation}
where the gauge coupling $g_{4}$ is equal to the strong coupling constant evaluated at the quark-lepton unification scale, and $(\nu_R)^c=(\nu^c)_L$. See Appendix~\ref{interactions} for details of the interactions in the physical basis.

The Yukawa interactions in Eq.~\eqref{eq:Yukawa}, other than generating the mass of the fermions, contain new Yukawa interactions with respect to the Standard Model. Particularly, the predicted scalar leptoquarks, $\Phi_3 \sim (\mathbf{\bar{3}}, \mathbf{2},-1/6)_\text{SM}$ and $\Phi_4 \sim (\mathbf{3}, \mathbf{2},7/6)_\text{SM}$, have the following interactions with quarks and leptons,
\begin{equation}
-{\cal L}_{QL}^Y  \supset  Y_2 \, Q_L \Phi_3 (\nu^c)_L  + Y_2 \, \ell_L  \Phi_4 (u^c)_L \ + Y_4 \, \ell_L  \Phi_3^\dagger (d^c)_L + \  Y_4 \, Q_L  \Phi_4^\dagger (e^c)_L +  {\rm h.c.}\, . 
\end{equation}
We note that neutrino masses can be small even when $Y_2 \to 0$, due to the inverse seesaw mechanism, but the entries in $Y_4$ cannot be arbitrarily small because one needs a realistic relation between down quarks and charged lepton masses,
\begin{equation}
Y_4=\sqrt{\frac{3}{2}} \, \frac{1}{v_2} (M_D - M_E).
\end{equation}
The scalar leptoquarks $\Phi_3$ and $\Phi_4$ can be written in $\SU(2)_L$ components as,
\beq
\Phi_3 =  \mqty(\phi_3^{1/3} \\[1ex]\phi_3^{-2/3} ), \hspace{0.5cm} \text{and} \hspace{0.5cm} \Phi_4 =\mqty(\phi_4^{5/3} \\[1ex]\phi_4^{2/3} ),
\eeq
where the numbers in the superscript denote the electric charge. In Appendix~\ref{interactions} we present the interactions of the leptoquarks in the physical basis, where the fermions are mass eigenstates. For some phenomenological studies of this theory see for example Refs.~\cite{Smirnov:1995jq,Faber:2018qon}.

\section{MESON DECAYS: $\mathcal{R}_K$ AND $\mathcal{R}_{K^*}$}
\label{sec:Bdecays}
%
The theory predicts the existence of a vector leptoquark, $X_\mu$, 
and two scalar leptoquarks, $\Phi_3$ and $\Phi_4$, among other fields. The interactions of the $X_\mu$ leptoquark are determined by several unknown unitary mixing matrices, see Appendix~\ref{interactions} for details. 
Unfortunately, one cannot explain easily the values 
of ${\cal{R}}_K$ and satisfy the bounds from the experimental constraints on $K_L \to e^{\pm} \mu^{\mp}$ when the mixing matrix is unitary. See the studies in Refs.~\cite{Faber:2018qon, Cornella:2021sby} for details. Consequently, we focus this study on the scalar leptoquarks that the theory predicts.

The interactions of the scalar leptoquarks with the Standard Model fermions are needed to render the fermion masses realistic, and therefore cannot be assumed small. 
In particular, in Eqs.~(\ref{eq:fermionmasses1}) and~(\ref{eq:fermionmasses2}) one can see that $Y_2$ can be neglected, but $Y_4$ must be non-zero in order to have a consistent relation between the charged leptons and down quarks masses.
For simplicity, in this section we will study scenarios where we take the limit $Y_2 \to 0$, and hence, the interactions of the scalar leptoquarks with the SM fermions are given by
\begin{equation}
-{\cal L}_Y =  Y_4^{ab}\left( \bar d_R^b (\phi_3^{1/3})^* \nu_L^a + \bar d_R^b (\phi_3^{-2/3})^* e_L^a +\bar e_R^b (\phi_4^{5/3})^* u_L^a +  \bar e_R^b (\phi_4^{2/3})^* d_L^a \right)+ \text{h.c.},
\end{equation}
which in the basis where the fermions are mass eigenstates read,
\begin{equation}
\begin{split}
-{\cal L}_Y = &  \ \bar d_R^b V_4^{ab}(\phi_3^{1/3})^* \nu_L^a + \bar d_R^b (K_3^* V_\text{PMNS}^* V_4)^{ab} (\phi_3^{-2/3})^* e_L^a  \\
&+ \bar e_R^b V_6^{ab} (\phi_4^{5/3})^* u_L^a +  \bar e_R^b (K_2 V_{\rm CKM}^T K_1 V_6)^{ab} (\phi_4^{2/3})^* d_L^a + \text{h.c.}\, .
\end{split}
\label{eq:YukawaInteractions}
\end{equation}

To determine the parameters quantifying the leptoquark interaction with fermions, i.e. the corresponding coupling and the leptoquark mass, the predictions of the theory should be contrasted with experimental measurements. Strikingly, both scalar leptoquarks contribute to $b \to s$ transitions through their coupling between the charged leptons and down quarks, via the $\phi_3^{-2/3}$ and $\phi_4^{2/3}$ fields. Therefore, in light of recent deviations reported by the LHCb on such transitions, we should ask the theory to accommodate the experimental results, being the largest deviation $3.1\sigma$ in the clean observable ${\cal{R}}_{K}$. In the following phenomenological analysis we are only considering the clean observables; namely, the ratios ${\cal R}_{K^{(*)}}$ and the branching fraction of the leptonic decays $\text{Br}(B_s \to \ell^+ \ell^-)$. 
\subsection{ Scalar Leptoquark \boldmath $ \phi_3^{-2/3}$} 
\label{sec:Phi3}
The scalar leptoquark  $\phi_3^{-2/3}$ contributes to the following dimension 6 effective interactions 
\begin{align}
{\cal L}_\text{eff}^{\phi_3^{-2/3}} \supset \frac{4G_F}{\sqrt{2}} V_{tb}V_{ts}^* \frac{\alpha}{4\pi}\left[ C_{9\ell \ell}' \, (\bar{s} \gamma_\mu P_R b ) (\bar{\ell} \gamma^\mu \ell) + C_{10 \ell \ell}' \,  (\bar{s} \gamma_\mu P_R b ) (\bar{\ell} \gamma^\mu \gamma^5 \ell)\right] + \text{h.c.},
\end{align}
whose Wilson coefficients are determined after integrating $\phi_3^{-2/3}$ out and are given by
\begin{align}
C_{10\ell \ell}' = - C_{9 \ell \ell}' & = \left(\frac{\sqrt{2} \pi }{G_F \, V_{tb}V_{ts}^* \, \alpha}\right)\frac{\left(K_3^* V_{\rm PMNS}^* V_4 \right)^{\ell 3} \left( K_3 V_{\rm PMNS} V_4^* \right)^{ \ell 2}}{4 M^2_{\phi_3^{-2/3}}} \nonumber\\[1.5ex]
& \simeq \left(36 \,\, {\rm TeV} \right)^2 \, \frac{\left(K_3^* V_{\rm PMNS}^* V_4 \right)^{\ell 3} \left( K_3 V_{\rm PMNS} V_4^* \right)^{ \ell 2}}{4 M^2_{\phi_3^{-2/3}}}. \label{eq:WilsonsPhi3}
\end{align}

The fact that the theory predicts $C_{10\ell \ell}'= -C_{9 \ell \ell}'$ allows us to write the leptonic branching ratio $B_s \to \ell^+ \ell^-$ as a function of a single Wilson coefficient $C_{10\ell \ell}'$,
\begin{equation}
\text{Br} (B_s \to \ell^+ \ell^- ) = \text{Br} (B_s \to \ell \ell )_\text{SM}  \times \left( 1 +0.4875 \, \text{Re}[ C_{10\ell \ell}' ] +  0.05940 \, |C_{10\ell \ell}'|^2 \right),
\end{equation}
where for the Standard Model prediction we take ${\rm Br}(B_s \to \mu^+ \mu^-)_{\rm SM} = 3.66 \times 10^{-9}$~\cite{Beneke:2019slt} and ${\rm Br}(B_s \to e^+ e^-)_{\rm SM} = 8.35 \times 10^{-14}$~\cite{Bobeth:2013uxa,Fleischer:2017ltw}. 
The same applies to the rest of the clean observables we are considering\footnote{For the calculation of the ratios ${\cal R}_{K}$ and ${\cal R}_{K^*}$ we adopt the form factors from Ref.~\cite{Bailey:2015dka} and Ref.~\cite{Straub:2015ica}, respectively.},
\begin{eqnarray}
{\cal R}_K &=& {\cal R}_{K}^\text{SM}\, \frac{1 - 0.5040\, \text{Re}[ C_{10\mu \mu}'] + 0.06359 |C_{10\mu \mu}'|^2}{1 - 0.5040 \, \text{Re}[C_{10ee}'] + 0.06359 |C_{10ee}'|^2}  \quad \quad \text{ for } q^2\subset [1.1, \, 6] \, \text{ GeV}^2,\\[1.5ex]
{\cal R}_{K^*} &=& {\cal R}_{K^*}^\text{SM} \, \frac{1 + 0.4335 \, \text{Re}[C_{10\mu \mu}'] + 0.07473 |C_{10\mu \mu}'|^2}{1 + 0.4325 \, \text{Re}[C_{10ee}'] + 0.07472 |C_{10ee}'|^2}  \quad \quad \text{ for }q^2\subset [1.1, \, 6] \, \text{GeV}^2,\\[1.5ex]
{\cal R}_{K^*}  &=& {\cal R}_{K^*}^\text{SM} \, \frac{1 + 0.2363 \, \text{Re} [C_{10 \mu \mu}'] + 0.03266 |C_{10\mu \mu}'|^2}{1 + 0.2252 \, \text{Re}[C_{10ee}'] + 0.03127 |C_{10ee}'|^2} \quad \quad \text{ for } q^2\subset [0.045, \, 1.1] \, \text{GeV}^2.
\end{eqnarray}
For the contributions to the Wilson coefficients from the SM we take $C_7^{\rm SM}=-0.304$, $C_9^{\rm SM} = 4.211$ and $C_{10}^{\rm SM} = -4.103$~\cite{Altmannshofer:2008dz}.

\begin{figure}[t]
\centering
\includegraphics[width=0.5\linewidth]{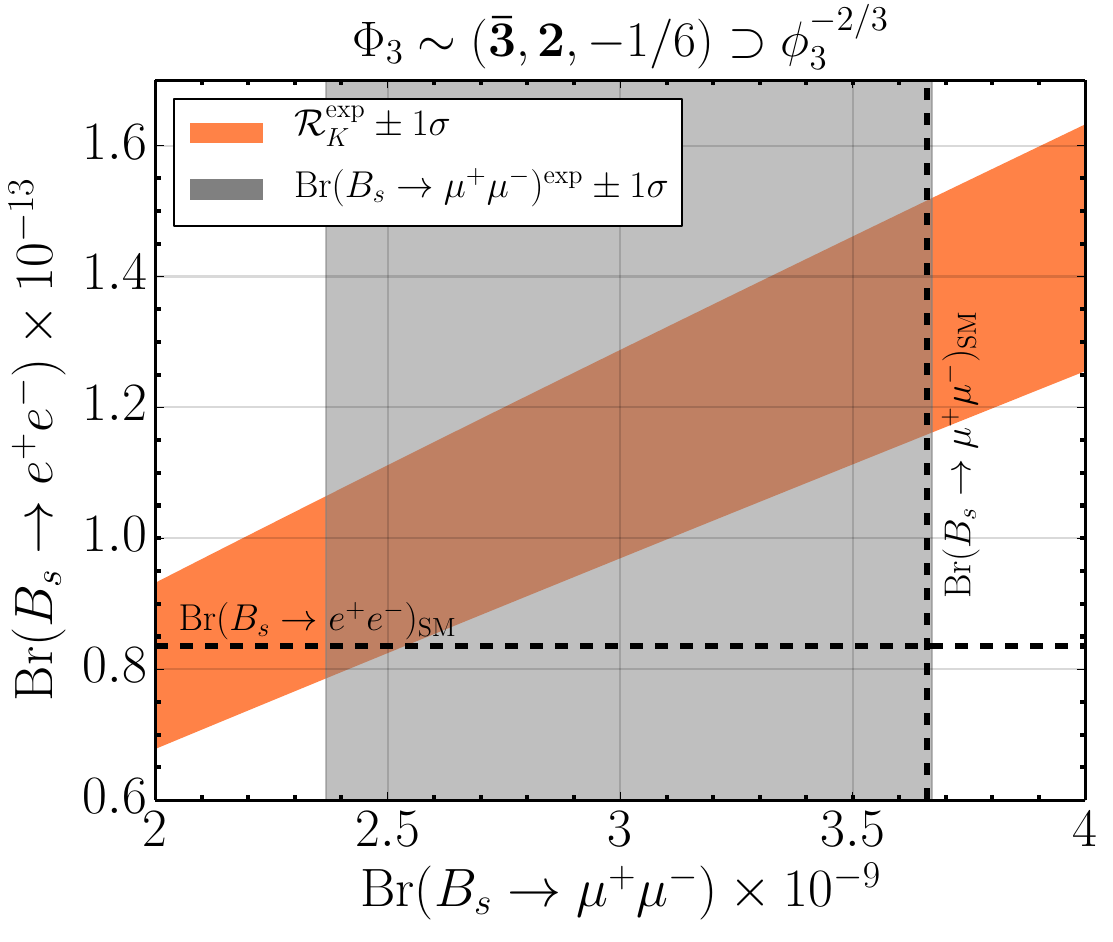}
\caption{The orange band gives the correlation between ${\rm Br}(B_s \to e^+ e^-)$ and ${\rm Br}(B_s \to \mu^+ \mu^-)$ that explains the $\mathcal{R}_K$ experimental measurement within 1$\sigma$. The gray band corresponds to the measurement of ${\rm Br}(B_s \to \mu^+ \mu^-)$ within $1\sigma$. The black dashed lines correspond to the SM predictions for each channel.
}
\label{fig:RKsatphi3}
\end{figure}

In Fig.~\ref{fig:RKsatphi3} we show the parameter space in the $\text{Br}(B_s \to \mu^+ \mu^-)$ vs $\text{Br}(B_s \to e^+ e^-)$ plane that satisfies the experimental value of ${\cal R}_K$ at the 1$\sigma$ level~\cite{Aaij:2021vac}, see Eq.~\eqref{eq:RKexp}.
We note that, due to the quadratic dependence of the observables on $C_{10\ell \ell}'$, for a given $\text{Br}(B_s \to \mu^+ \mu^-)$ and $\text{Br}(B_s \to e^+ e^-)$ there exist four possible values of ${\cal R}_K$ allowed. In this figure we present the solution that is also consistent with the measured window for ${\cal R}_K$. We also take into account the existing experimental bounds on the leptonic decay to muons~\cite{Aaij:2017vad} and to electrons~\cite{Aaltonen:2009vr}, which are given by
\begin{align}
\text{Br} ( B_s \to \mu^+ \mu^-)^\text{exp} & = 3.0 \pm 0.6^{+0.3}_{-0.2} \times 10^{-9},\\[1ex]
\text{Br}(B_s \to e^+ e^-)^\text{exp} & < 2.8 \times 10^{-7}.
\end{align}
The region shaded in gray in Fig.~\ref{fig:RKsatphi3} shows explicitly the parameter space satisfying $\text{Br}(B_s \to \mu^+ \mu^-)^\text{exp}$~\cite{Aaij:2017vad} at the 1$\sigma$ level. As the figure shows, we find that there is a region of the parameter space that satisfies ${\cal R}_K^\text{exp}$ and $\text{Br}(B_s \to \mu^+ \mu^-)^\text{exp}$ at 1$\sigma$ which corresponds to the overlapping region between the regions shaded in gray and in orange in the plot.

\begin{figure}[b]
\centering
\includegraphics[width=0.45\linewidth]{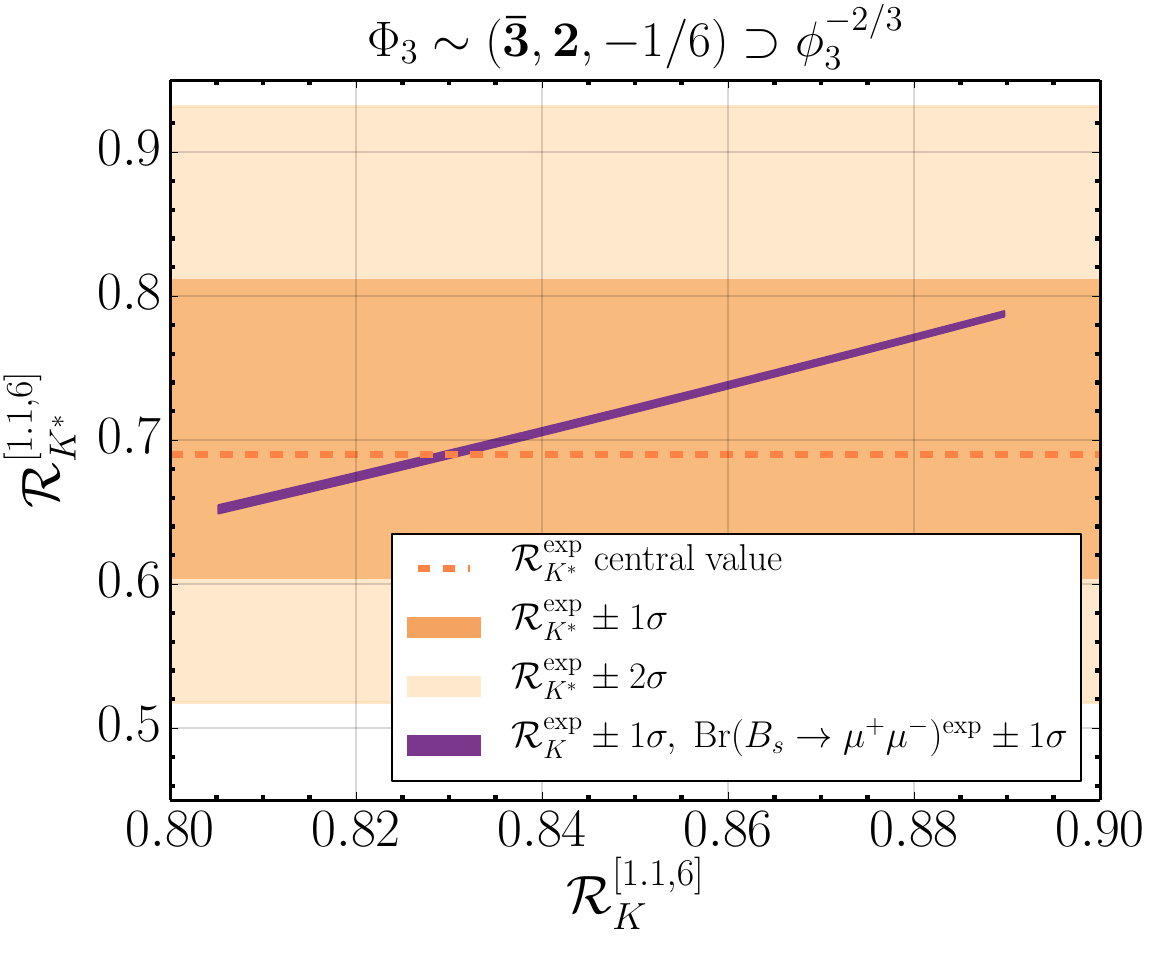}
\includegraphics[width=0.45\linewidth]{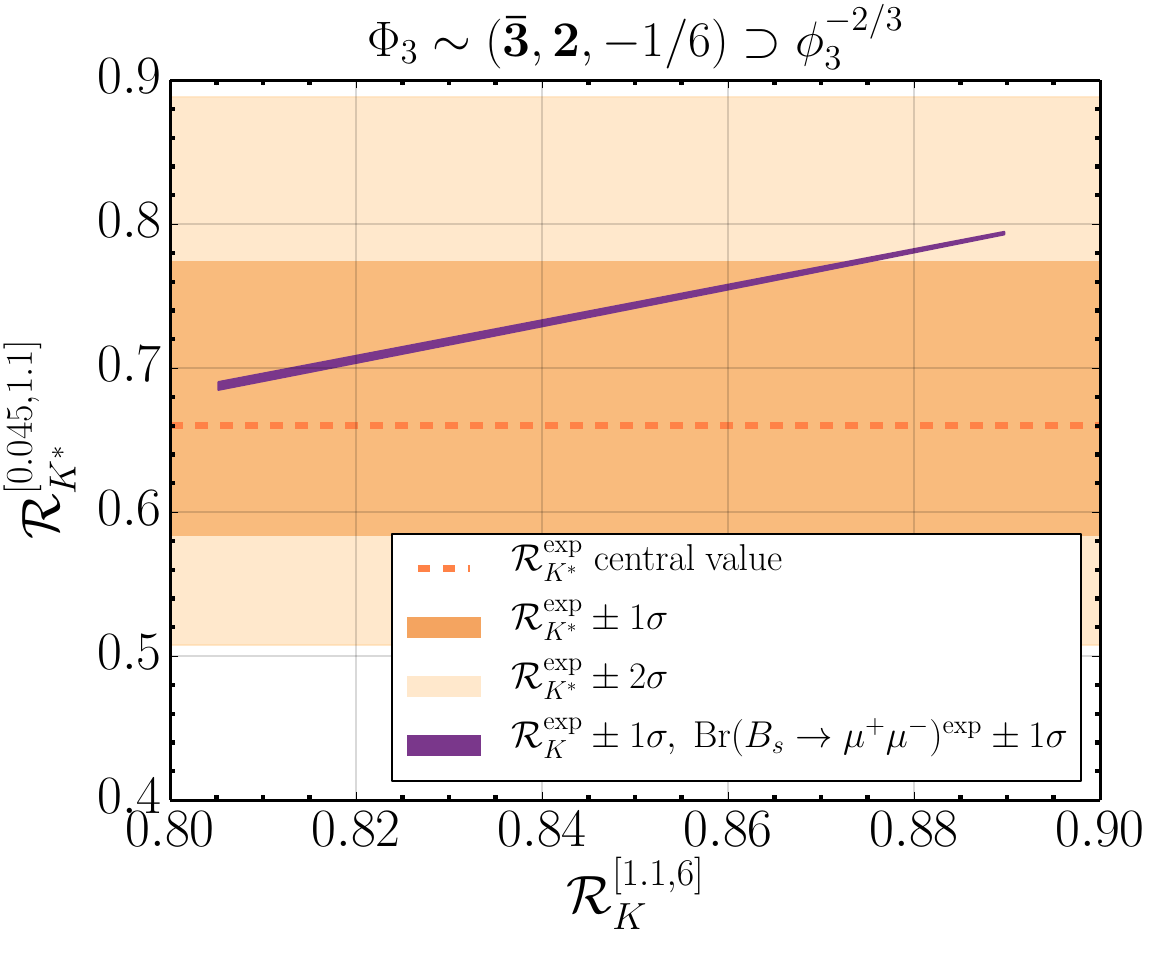}
\caption{The purple band gives the prediction for $\mathcal{R}_{K^*}$ in the window $0.045< q^2 <1.1 \,\, \text{ GeV}^2$  (left panel) and $1.1<q^2<6 \,\, \text{GeV}^2$ (right panel) for the points satisfying ${\cal R}_K^\text{exp}$ and $\text{Br}(B_s \to \mu^+ \mu^-)^\text{exp}$ within $1\sigma$. The region shaded in orange (light orange) corresponds to the measurement of ${\cal R}_{K^*}$ at $1\sigma$ ($2\sigma$).}
\label{fig:RKRKstarphi3}
\end{figure}

In Fig.~\ref{fig:RKRKstarphi3} we plot the correlation between the semileptonic ratios for the different $q^2$ ranges tested at experiment for the values of $\text{Br}(B_s \to \mu^+ \mu^-)$ and $\text{Br}(B_s \to e^+ e^- )$ consistent with the experimental values of ${\cal R}_K^\text{exp}$ and $\text{Br}(B_s \to \mu^+ \mu^-)^\text{exp}$ at the 1$\sigma$ level. We note that the theory predicts a window for ${\cal R}_{K^*}$ that is consistent with the experimental values of this observable~\cite{Aaij:2017vbb}, 
\begin{equation}
\label{eq:RKstarEXP}
    {\cal R}_{K^*}^\text{exp} = \begin{cases}
    0.66^{+0.11}_{-0.07} \text{ (stat) }\pm 0.03 \text{ (syst) }  & \text{ for } 0.045 < q^2 < 1.1 \text{ GeV}^2,\\[1.5ex]
    0.69^{+0.11}_{-0.07} \text{ (stat) } \pm 0.05 \text{ (syst) } & \text{ for } 1.1<q^2<6.0 \text{ GeV}^2,
  \end{cases}
\end{equation}
which deviate from the Standard Model prediction by $2.2\sigma$ and $2.4 \sigma$, respectively. Since the main focus of our work is to explain $\mathcal{R}_K$ and $\text{Br}(B_s \to \mu^+ \mu^-)$ we aim to reproduce these observables within $1\sigma$, while for $\mathcal{R}_{K^*}$ we consider the $2\sigma$ ranges. As Fig.~\ref{fig:RKRKstarphi3} shows, for $\mathcal{R}_{K^*}$ in the range $1.1 < q^2 < 6 \text{ GeV}^2$ the full predicted window is in agreement with the experimental measurement within $1\sigma$, while for $0.045 < q^2 < 1.1 \text{ GeV}^2$, the theory prefers higher values still within $2\sigma$ of its experimental value. 

\begin{figure}[t]
\centering
\includegraphics[width=0.5\linewidth]{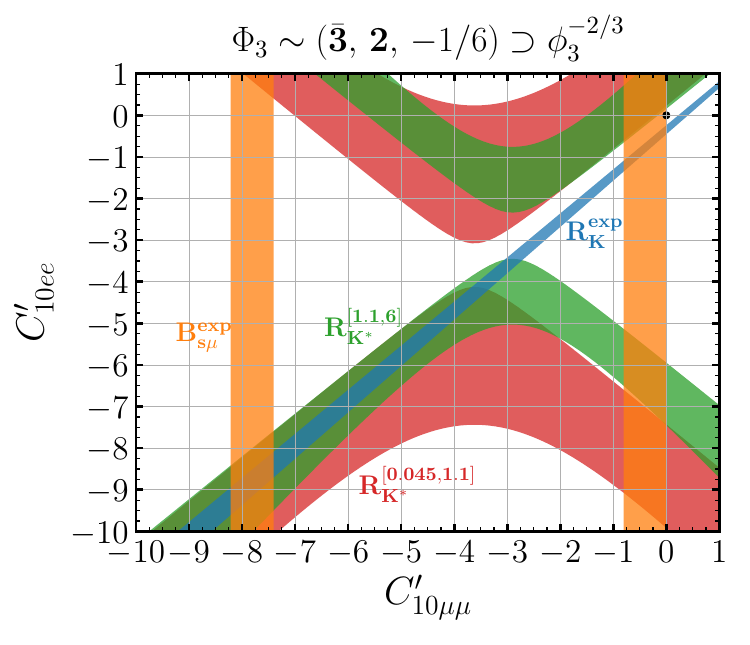}
\caption{
Parameter space of the Wilson coefficients $C'_{10ee}$ and $C'_{10\mu \mu}$ required to explain the flavor anomalies. The orange band is in agreement with the measurement of ${\rm Br}(B_s \to \mu^+ \mu^-)$ and the blue band with $\mathcal{R}_K$ within $1\sigma$. The red and green bands correspond to the measurements of $\mathcal{R}_{K^*}^{[0.045,1.1]}$ and $\mathcal{R}_{K^*}^{[1.1,6]}$  within $2\sigma$, respectively. }
\label{fig:ParameterSpacePhi3}
\end{figure}

Fig.~\ref{fig:ParameterSpacePhi3} shows the parameter space in the plane of the relevant Wilson coefficients $C'_{10 \mu \mu}$ and $C'_{10ee}$ that satisfies the experimental value of the clean observables at the $1\sigma$ level: $\text{Br}(B_s \to \mu^+ \mu^-)^\text{exp}$ in orange and ${\cal R}_K^\text{exp}$ in blue. For  $\mathcal{R}_{K^*}^{[0.045,1.1]}$ and $\mathcal{R}_{K^*}^{[1.1,6]}$ we consider the $2\sigma$ range of the measurement which are shown in red and green, respectively; between brackets we specify the window of the integrated $q^2$. As can be seen from the lower left part of the plot, all observables can be explained by Wilson coefficients $C'_{10\mu\mu}\approx -8$ and $C'_{10ee}\approx -9$.

As we have shown in this subsection, the simplest theory where one can understand unification of matter at the TeV scale naturally accommodates the so-called flavor anomalies in $b \to s$ transitions observed at experiment. As one can read from the Wilson coefficients in Eq.~\eqref{eq:WilsonsPhi3}, such transitions particularly imply the presence of four entries in the Yukawa matrix between the charged leptons and the down quarks, which in the physical basis reads as $\tilde V_4 \equiv K_3^* V_\text{PMNS}^* V_4$, as can be read from Eq.~\eqref{eq:YukawaInteractions}. However, the rest of the couplings in this matrix may lead to other flavor transitions, such as $K_L \to \mu^\pm e^\mp$ and $\tau$ decays to light mesons and a charged lepton, which suffer from strong experimental constraints. Requiring consistency with the experiment allows us to infer the texture of the Yukawa matrix $\tilde V_4$. By adopting the following hierarchy in their entries,
\begin{equation}
\tilde V_4  =
\begin{pmatrix}
\phantom{a} \cdot \phantom{a} &\phantom{a}  \hlight{\phantom{a}} \phantom{a} &\phantom{a}  \hlight{\phantom{a}} \phantom{a}  \\ 
\phantom{a}  \cdot \phantom{a} &\phantom{a}  \hlight{\phantom{a}}\phantom{a}  &\phantom{a}  \hlight{\phantom{a}} \phantom{a} \\
\phantom{a}  \cdot \phantom{a} &\phantom{a}  \cdot \phantom{a}  &\phantom{a} \cdot \phantom{a} 
\end{pmatrix},
\end{equation}
where the squares denote large entries while the dots denote small entries, the theory can accommodate the experimental values of the clean observables involving $b \to s$ transitions while being consistent with all existing flavor constraints. Notice that lepton flavor violation processes such as the radiative decays $\mu \to e \gamma$ or the muon magnetic moment $(g-2)_\mu$ do not offer relevant constraints to the four matrix entries involved in $b \to s$ transitions since they suffer from the muon mass suppression and a near cancellation of the loop functions due to the leptoquark charge of $|Q_{\rm LQ}|=2/3$. We refer the reader to Appendix~\ref{appMEG} for more details. 

The leptoquark $\phi_3^{-2/3}$ can also induce the decays $B \to K \mu^\pm e^\mp$ and $B_s\to \mu^\pm e^\mp$; however, these decays depend on a different combination of couplings than the ones that enter in $\mathcal{R}_{K^{(*)}}$, and hence, there exists enough freedom in the Wilson coefficients for this decay to satisfy the experimental constraint. Furthermore, $C'_{10}$ will also modify the $B_s - \bar{B}_s$ mass difference at one-loop that leads to the constraint $|C'_{10}| M_{\phi_3^{2/3}} \lesssim 100$ TeV~\cite{Becirevic:2015asa}. Since we need $C'_{10} \simeq -9$ to explain the flavor anomalies, the mass of the leptoquark must satisfy $M_{\phi_3^{2/3}} \lesssim 10$ TeV. 

On the other hand, we note that $\phi_3^{1/3}$, which also belongs to the $\SU(2)_L$ doublet  in Eq.~\eqref{eq:YukawaInteractions}, shares the entries of $\tilde V_4$ up to the effect of the $V_\text{PMNS}$ and some complex diagonal phases. Knowing the texture of $\tilde V_4$ from the interactions involving $\phi_3^{-2/3}$, the theory predicts a modification of processes such as $B \to K^{(*)} \nu \bar{\nu} $ and $B_s \to \nu \bar{\nu}$. The current experimental bound on the former is given by ${\rm Br} (B^+ \to K^{+} \nu \bar{\nu} )< 1.7 \times 10^{-5}$~\cite{Lees:2013kla} which is a factor of three larger than the prediction in the SM ${\rm Br} (B^+ \to K^{+} \nu \bar{\nu} )_{\rm SM} = 5.6\times 10^{-6}$~\cite{Kou:2018nap}, and ${\rm Br} (B^+ \to K^{*+} \nu \bar{\nu} )< 4 \times 10^{-5}$~\cite{Lutz:2013ftz}, still far from the SM prediction ${\rm Br} (B^+ \to K^{*+} \nu \bar{\nu} )_{\rm SM} = 9.6\times 10^{-6}$~\cite{Kou:2018nap}. Because of this and the large uncertainties in the hadronic form factors it is hard to obtain constraints from these observables.

\FloatBarrier

\subsection{Scalar Leptoquark \boldmath $ \phi_4^{2/3}$}
\label{sec:Phi4}
The scalar leptoquark $\phi_4^{2/3}$ contributes to the following dimension 6 effective interactions,
\begin{equation}
    {\cal L}_\text{eff}^{\phi_4^{2/3}} = \frac{4G_F}{\sqrt{2}}  V_{tb}V_{ts}^* \frac{\alpha}{4\pi} \left[C_{9\ell \ell}(\bar s \gamma_\mu P_L b)(\bar \ell \gamma^\mu \ell) + C_{10\ell \ell}(\bar s \gamma_\mu P_L b)(\bar \ell \gamma^\mu \gamma_5 \ell)  \right] + \text{h.c.},
\end{equation}
where the Wilson coefficients are given by
\begin{equation}
    C_{10 \ell \ell} = C_{9 \ell \ell} = -\left(\frac{\pi \sqrt{2}}{G_F \, V_{tb}V_{ts}^* \, \alpha}\right) \frac{(K_2 V_\text{CKM}^T K_1 V_6)^{3\ell}(K_2^* V_\text{CKM}^\dagger K_1^* V_6^*)^{2\ell}}{4 M_{\phi_4^{2/3}}^2}.
\end{equation}
In this case the leptonic branching ratio is also given as a function of a single Wilson coefficient, $C_{10 \ell \ell}$,
\begin{equation}
  \text{Br}(B_s \to \ell^+ \ell^-) = \text{Br}( B_s \to \ell^+ \ell^-)_\text{SM} \times \left(  1 - 0.487448 \, \text{Re} \, [ C_{10\ell \ell} ] + 0.0594014 |C_{10\ell \ell}|^2 \right),
\end{equation}
as well as the other clean observables we consider,
\begin{eqnarray}
{\cal R}_K &=& {\cal R}_K^\text{SM}  \frac{1 - 0.01812 \, \text{Re} [C_{10\mu \mu}] + 0.06359 |C_{10\mu \mu}|^2}{1 - 0.01781 \, \text{Re}[C_{10ee}] + 0.06359 |C_{10ee}|^2} \quad \quad \text{ for } q^2\subset [1.1, \, 6] \, \text{ GeV}^2,\\[1.5ex]
{\cal R}_{K^*} &=& {\cal R}_{K^*}^\text{SM}  \frac{1 - 0.08301 \, \text{Re}[C_{10\mu \mu} ] + 0.07473 |C_{10\mu \mu}|^2}{1 - 0.08428 \, \text{Re}[ C_{10ee} ] + 0.07472 |C_{10ee}|^2} \quad \quad \text{ for } q^2\subset [1.1, \,6] \, \text{ GeV}^2,\\[1.5ex]
{\cal R}_{K^*} &=& {\cal R}_{K^*}^\text{SM}  \frac{1 - 0.04783 \, \text{Re}[ C_{10\mu \mu} ] + 0.03266 |C_{10\mu \mu}|^2}{1 - 0.04600 \, \text{Re}[C_{10ee}] + 0.03127 |C_{10ee}|^2} \quad \quad    \text{ for } q^2\subset [0.045, \, 1.1] \, \text{ GeV}^2.
\end{eqnarray}

\begin{figure}[t]
\centering
\includegraphics[width=0.47\linewidth]{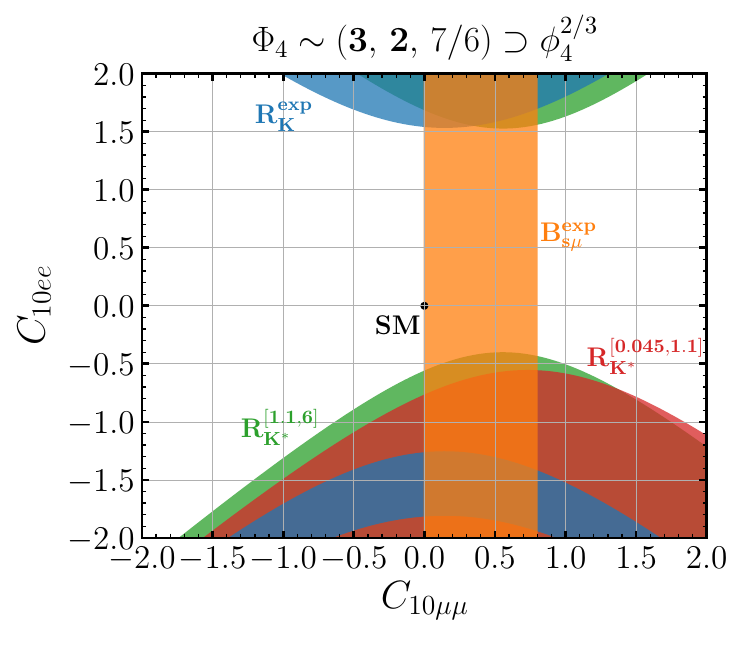}
\includegraphics[width=0.46\linewidth]{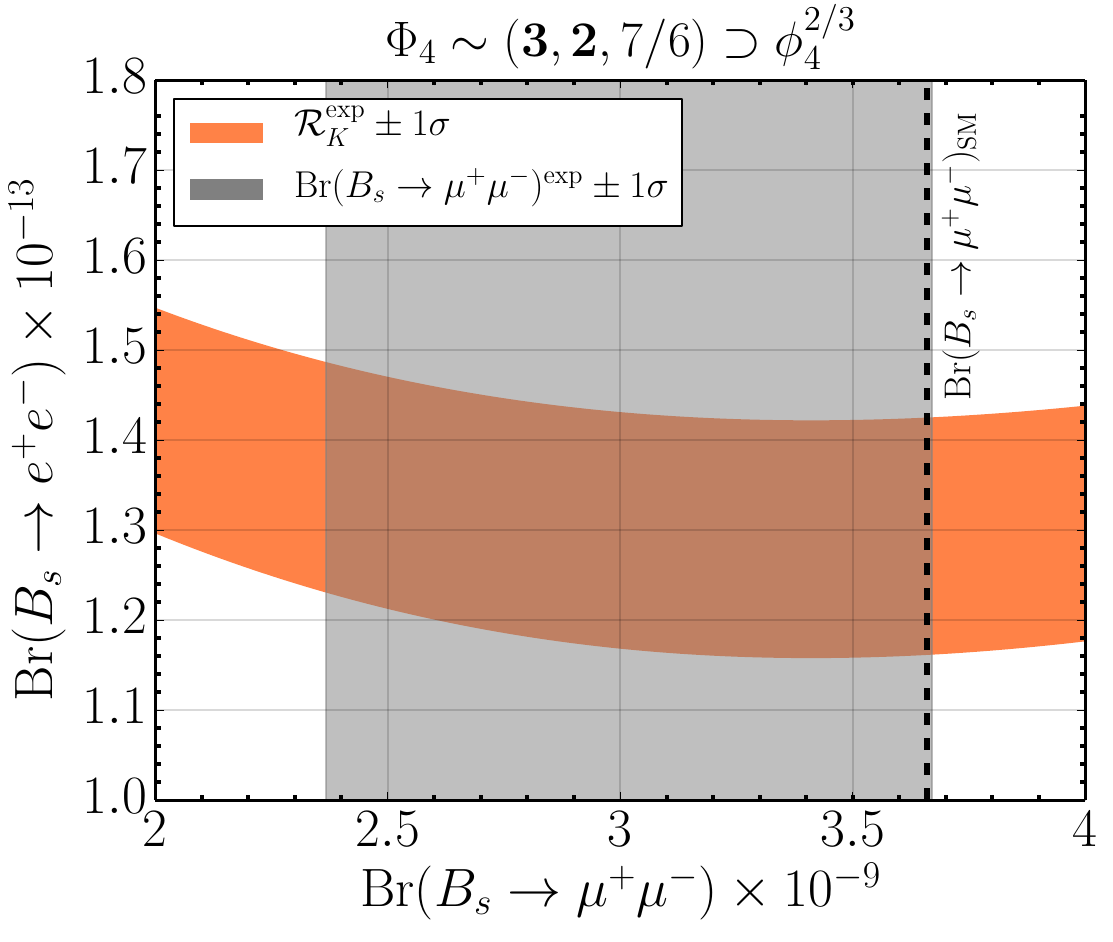}
\caption{ \textit{Left panel:} Parameter space of the Wilson coefficients $C_{10ee}$ and $C_{10\mu \mu}$ required to explain the flavor anomalies. The orange band is in agreement with the measurement of ${\rm Br}(B_s \to \mu^+ \mu^-)$ and the blue band with $\mathcal{R}_K$ within $1\sigma$. The red and green bands correspond to the measurements of $\mathcal{R}_{K^*}^{[0.045,1.1]}$ and $\mathcal{R}_{K^*}^{[1.1,6]}$  within $2\sigma$, respectively. \textit{Right panel:} The orange band gives the correlation between ${\rm Br}(B_s \to e^+ e^-)$ and ${\rm Br}(B_s \to \mu^+ \mu^-)$ that explains the $\mathcal{R}_K$ experimental measurement within 1$\sigma$. The gray band corresponds to the measurement of ${\rm Br}(B_s \to \mu^+ \mu^-)$ within $1\sigma$. In both panels, only the region where $C_{10\mu \mu}$ is very close to zero is allowed by the $\mu \to e \gamma$ constraint. Therefore, the prediction for $\text{Br}(B_s \to \mu^+ \mu^-)$ is very close to the SM prediction; the latter is shown by a black dashed line.}
\label{fig:ParameterSpacePhi4}
\label{RKRKstarphi4}
\end{figure}

In the left panel in Fig.~\ref{fig:ParameterSpacePhi4} we show our results in the plane of the Wilson coefficients $C_{10ee}$ vs $C_{10\mu\mu}$. The orange band reproduces the measured value for $\text{Br}(B_s \to \mu^+ \mu^-)$ while the blue band reproduces $\mathcal{R}_K$ within $1\sigma$. The red and green bands reproduce the measurements of $\mathcal{R}_{K^*}^{[0.045,1.1]}$ and $\mathcal{R}_{K^*}^{[1.1,6]}$ within $2\sigma$, respectively. In contrast to the $\Phi_3$ leptoquark, $\Phi_4$ can reproduce the observables with small Wilson coefficients and coupling mainly to electrons; namely, $C_{10\mu\mu} \approx 0$ and $C_{10ee}\approx -1.3$. Therefore they will have a smaller impact on each separate channel. Note that the component $\phi_4^{5/3}$ can give large contributions to $\mu \to e \gamma$ since the near cancellation on the loop functions that takes place for $\phi_4^{2/3}$ does not take place for $\phi_4^{5/3}$, see Appendix~\ref{appMEG} for more details. Therefore, in order to be consistent with the lepton flavor violation constraints, the following texture in the Yukawa matrix $\tilde V_6 = K_2 V_\text{CKM}^T K_1 V_6$ must be adopted,
\begin{equation}
\tilde V_6  =
\begin{pmatrix}
\phantom{a} \cdot \phantom{a} & \phantom{a}  \cdot \phantom{a} & \phantom{a}  \cdot \phantom{a}  \\ 
\phantom{a}  \hlight{\phantom{a}}\phantom{a}  & \phantom{a}  \cdot \phantom{a}  &\phantom{a}  \cdot \phantom{a} \\
\phantom{a}  \hlight{\phantom{a}}  \phantom{a} & \phantom{a} \cdot \phantom{a}  &\phantom{a} \cdot \phantom{a} 
\end{pmatrix},
\label{eq:Y4texturePhi4}
\end{equation}
where the leptoquark $\Phi_4$ couples mostly to electrons.

\begin{figure}[t]
\centering
\includegraphics[width=0.45\linewidth]{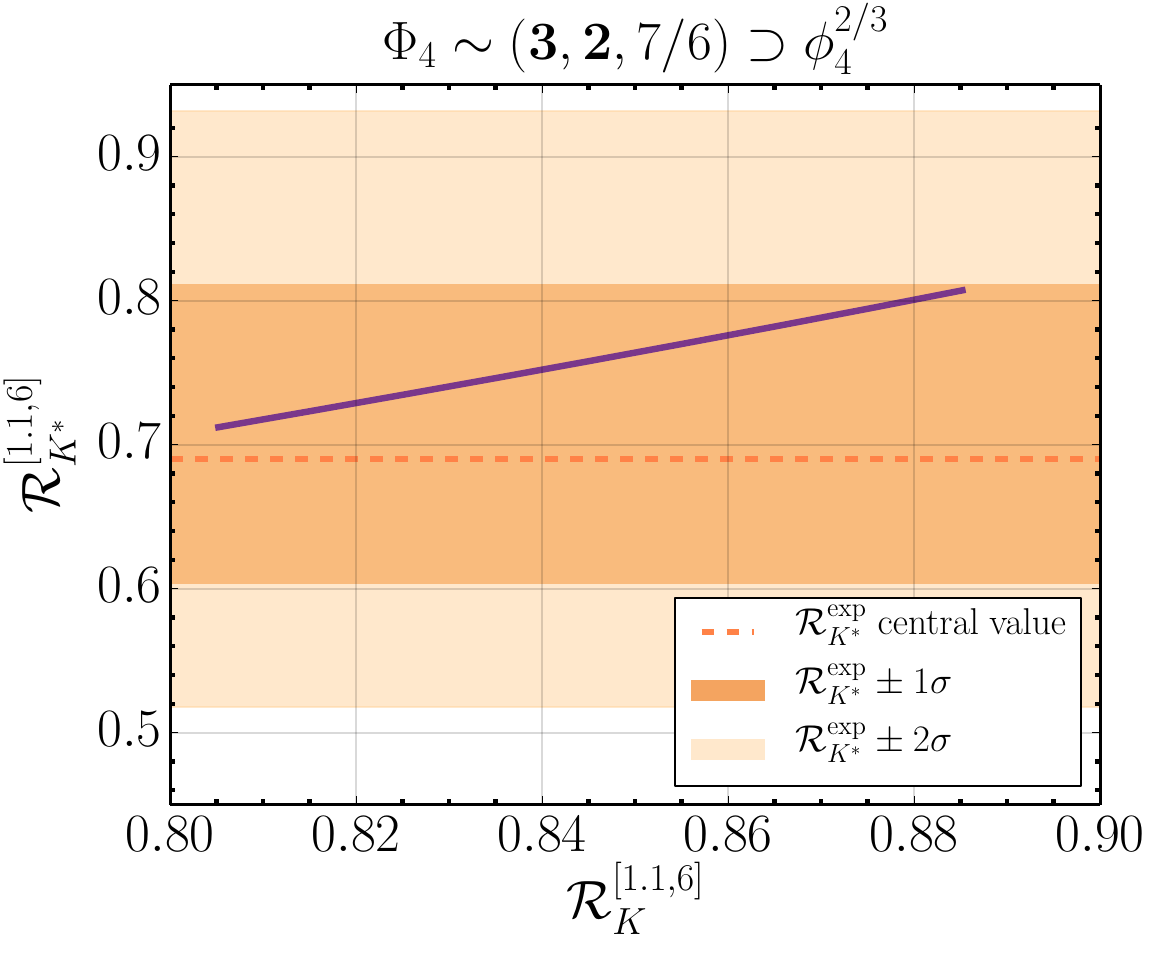}
\includegraphics[width=0.45\linewidth]{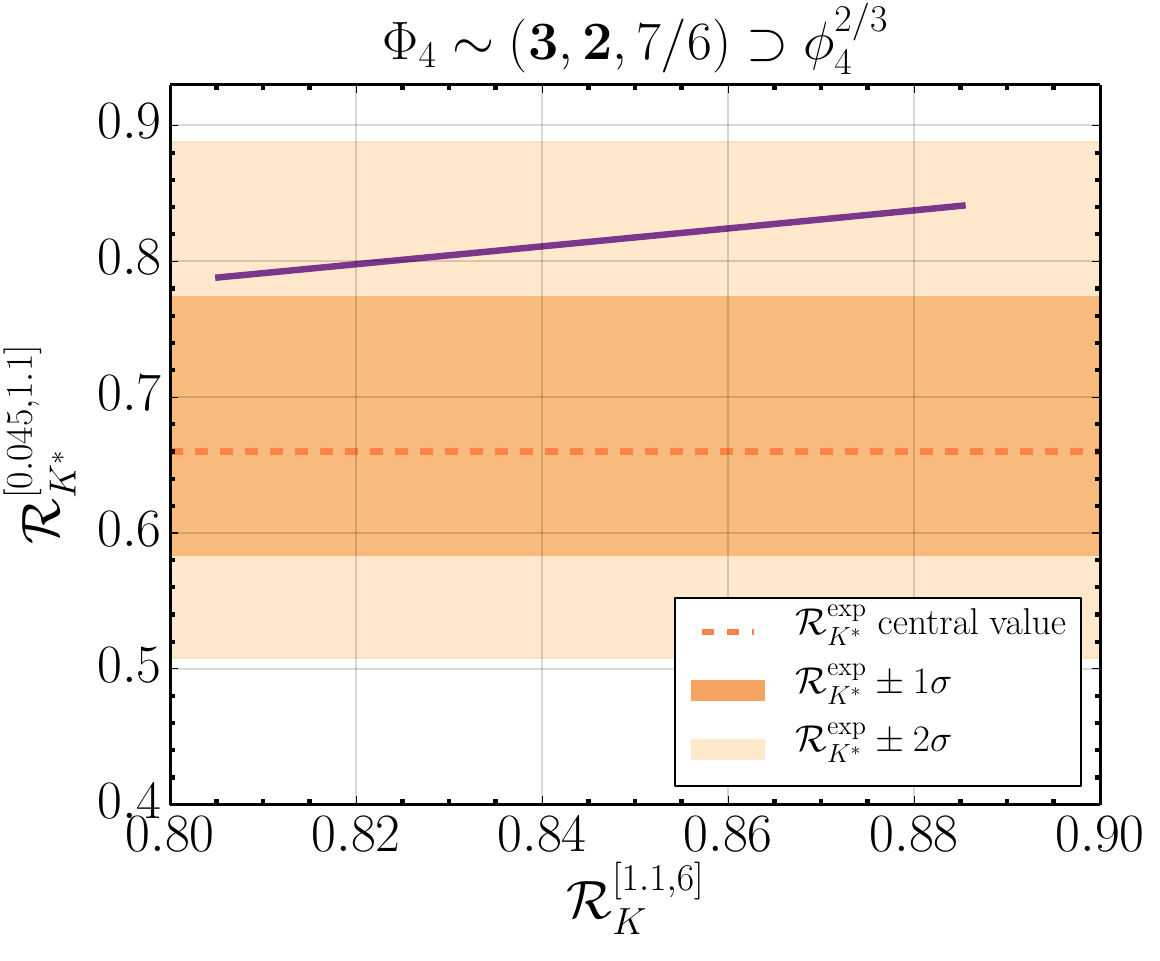}
\caption{The purple band gives the prediction for $\mathcal{R}_{K^*}$ in the window $0.045< q^2 <1.1 \,\, \text{ GeV}^2$  (left panel) and $1.1<q^2<6 \,\, \text{GeV}^2$ (right panel) for the points satisfying ${\cal R}_K^\text{exp}$ within $1\sigma$. The region shaded in orange (light orange) corresponds to the measurement of ${\cal R}_{K^*}$ at $1\sigma$ ($2\sigma$). We show the solution with small Wilson coefficients.}
\label{fig:RKRKstarphi4}
\end{figure}

In Fig.~\ref{fig:RKRKstarphi4} we show the correlation predicted for the ratios ${\cal R}_{K^*}$ and ${\cal R}_K$ adopting the texture in Eq.~\eqref{eq:Y4texturePhi4}. In the left panel we give the predictions\footnote{We note that the experimental value $\text{Br}(B_s \to \mu^+ \mu^-)^\text{exp}$ is already consistent with the SM prediction at $1\sigma$.} for ${\cal R}_{K^*}$ in the window $1.1<q^2<6 \,\, {\rm GeV}^2$, where the purple band is in agreement with ${\cal R}_K$ within $1\sigma$. In the right panel we show the predictions for ${\cal R}_{K^*}$ in the window $0.045<q^2<1.1 \,\, {\rm GeV}^2$.

\section{THE \boldmath $g-2$ OF THE MUON}
\label{sec:muong2}
%
\begin{figure}[b]
\centering
\includegraphics[width=0.75\linewidth]{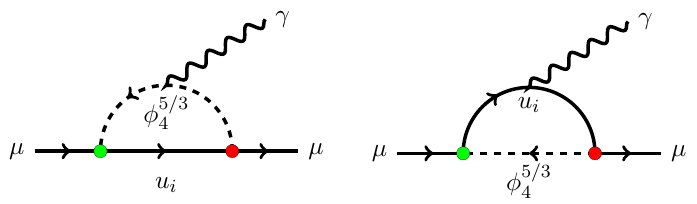}
\caption{Feynman diagrams for the topologies of the main contributions from the scalar leptoquark $\phi_4^{5/3}$ to $(g-2)_\mu$. The different colors in the vertices indicate opposite chiralities in the leptoquark - muon coupling.}
\label{fig:topologiesgm2}
\end{figure}
The Fermilab $g-2$ experiment has recently reported results on the anomalous magnetic moment of the muon $a_\mu\equiv (g-2)_\mu/2$ from their Run 1~\cite{Abi:2021gix}. The combined result with the one from the E821 experiment at BNL~\cite{Bennett:2006fi} deviates from the SM prediction by $4.2\sigma$, as Eq.~\eqref{eq:FermiLab} manifests.
In this section we show that in the most general case, the theory gives a prediction for $(g-2)_\mu$ that can explain the reported deviation, involving the same leptoquarks in the theory that have been discussed so far.\\

\begin{figure}[t]
\centering
\includegraphics[width=0.5\linewidth]{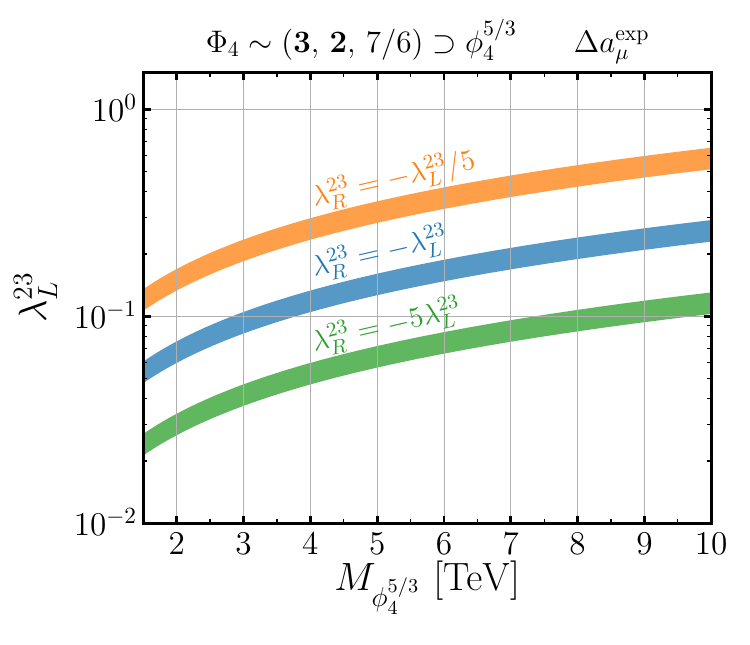}
\caption{Parameter space in the $\lambda_L^{23}$ vs $M_{\phi_4^{5/3}}$ plane, the shaded regions reproduce the combined result from the Muon $g−2$ experiment at Fermilab and E821 at BNL within $1\sigma$. Different colors correspond to different relations between the couplings $\lambda_L^{23}$ and $\lambda_R^{23}$.
}
\label{fig:muong21}
\end{figure}

In this theory, the main contribution to the muon $g-2$ is generated by the scalar leptoquark $\phi_{4}^{5/3}$ with the top quark running in the loop. In Fig.~\ref{fig:topologiesgm2} we show the Feynman graphs for the two different 
topologies. The relevant Yukawa interactions for the $\phi_{4}^{5/3}$ field are given by
\begin{equation}
- {\cal L} \supset  \,\, \bar e^i \left( \lambda_R^{ij} \, P_L + \lambda_L^{ij} \,\, P_R \right)  u^j \, \left(\phi_4^{5/3}\right)^* \ + \ \text{h.c.} \ ,
\end{equation}
where the matrices $\lambda_{L,R}$ correspond to 
\begin{equation}
\label{eq:lambdaLR}
\lambda_{R} = V_6^T  ,  \hspace{0.8cm} {\rm and} \hspace{0.8cm} \lambda_{L} = -K_3 V_{\rm PMNS} V_5^* ,
\end{equation}
and these matrices can be written in terms of the Yukawa matrices in the Lagrangian as  $V_5=N^T Y_2 U_C$ and $V_6=U^T Y_4 E_C$ as discussed in Appendix~\ref{interactions}. The new contribution to the muon $g-2$ can be written as
\begin{eqnarray}
\Delta a_{\mu} &=&  \frac{-3}{16 \pi^2} \frac{m_\mu^2}{M_{\phi_4^{5/3}}^2} \sum_{j} \left [ \left( |\lambda_{L}^{2j}|^2 + |\lambda_{R}^{2j}|^2 \right) \times \left(  \frac{2}{3} F_1 (x_j) + \frac{5}{3} F_2 (x_j)\right)  \nonumber \right. \\
&&\left. + \frac{m_{u_j}}{m_\mu} {\rm{Re}}[ \lambda_{L}^{2j} (\lambda_{R}^{2j})^* ] \left(\frac{2}{3} F_3 (x_j) + \frac{5}{3} F_4 (x_j) \right) \right] ,
\label{eq:amu}
\end{eqnarray}
where the loop-functions are given by
\begin{eqnarray}
F_1 (x)&=& \frac{1}{6(1-x)^4} (2 + 3 x - 6 x^2 + x^3 + 6 x \ {\rm{ln}} x ), \\
F_2 (x) &=& \frac{1}{6(1-x)^4} (1-6x+3 x^2 + 2 x^3 - 6 x^2 \ {\rm{ln}} x),\\
F_3 (x) &=& \frac{1}{(1-x)^3} (-3 + 4 x - x^2 - 2 \ {\rm{ln}}x), 
\end{eqnarray}
\begin{eqnarray}
F_4 (x) &=& \frac{1}{(1-x)^3} (1-x^2 + 2 x \ {\rm{ln}} x),
\end{eqnarray}
with $x_j=\left({m_{u_j}}/{M_{\phi_4^{5/3}}}\right)^2$. Therefore, the dominant contribution will come from the top quark in the loop.

We present our results in Fig.~\ref{fig:muong21} in the $\lambda_L^{23}$ vs $M_{\phi_4^{5/3}}$ plane. The shaded bands are in agreement with the combined result from the Muon $g−2$ experiment at Fermilab and E821 at BNL within $1\sigma$ given in Eq.~\eqref{eq:FermiLab}. The orange band corresponds to fixing $\lambda_R^{23}=-\lambda_L^{23}/5$, the blue band is for $\lambda_R^{23}=-\lambda_L^{23}$ while the green band is for $\lambda_R^{23}=-5\lambda_L^{23}$.

The contributions from $\phi_3^{-2/3}$ and $\phi_4^{2/3}$ to the muon $g-2$ have chiral suppression and although this can be lifted through mixing, the latter is determined by the electroweak scale and generically it is very small; furthermore, the constraints from the LHC rule out such scenario as an explanation for the $(g-2)_\mu$ anomaly~\cite{Dorsner:2019itg}.

\subsection{Connection between the Flavor and Muon $g-2$ Anomalies}
\label{all}
%
In this section we study the possibility to explain the $\mathcal{R}_K$ and the $(g-2)_\mu$ anomalies simultaneously. For alternative solutions to both of these anomalies see e.g. Refs.~\cite{Bauer:2015knc,Saad:2020ihm,Huang:2020ris,Babu:2020hun,Greljo:2021xmg,Marzocca:2021azj}. In the theory discussed in this work, the flavor and muon $g-2$ anomalies can be simultaneously  explained when we consider the contribution from both $\Phi_3$ and $\Phi_4$. As we have discussed, $\Phi_4$ can explain the $g-2$ of the muon and in order to avoid the strong experimental constraint on $\mu \to e \gamma$ we assume that $\Phi_4$ couples mainly to muons by adopting the following texture
\beq
\tilde{V}_6 = K_2 V^T_{\rm CKM} K_1 V_6 = \begin{pmatrix}
\phantom{a} \cdot \phantom{a} &\phantom{a}  \cdot \phantom{a} &\phantom{a}  \cdot \phantom{a}  \\ 
\phantom{a}  \cdot \phantom{a} &\phantom{a}  \hlight{\phantom{a}}\phantom{a}  &\phantom{a}  \cdot \phantom{a} \\
\phantom{a}  \cdot \phantom{a} &\phantom{a}  \hlight{\phantom{a}}  \phantom{a}  &\phantom{a} \cdot \phantom{a} 
\end{pmatrix}.
\label{eq:textureV4}
\eeq
As we studied in Section~\ref{sec:Phi3}, $\Phi_3$ can be used to explain the flavor anomalies by being coupled to both electrons and muons, and hence, we assume the following texture
\beq
\tilde{V}_4 = K_3^* V^*_{\rm PMNS} V_4 = \begin{pmatrix}
\phantom{a} \cdot \phantom{a} &\phantom{a}  \hlight{\phantom{a}}\phantom{a} &\phantom{a}  \hlight{\phantom{a}}\phantom{a}  \\ 
\phantom{a}  \cdot \phantom{a} &\phantom{a}  \hlight{\phantom{a}}\phantom{a}  &\phantom{a}  \hlight{\phantom{a}}\phantom{a} \\
\phantom{a}  \cdot \phantom{a} &\phantom{a}  \cdot \phantom{a} &\phantom{a} \cdot \phantom{a} 
\end{pmatrix}.
\eeq
As given in Eq.~\eqref{eq:lambdaLR} the coupling $\lambda_L^{23}$ is determined by the matrix $V_5$. Therefore, for the matrix $V_5$ we only require the entry $V_5^{23}$ to be non-zero.

\begin{figure}[t]
\centering
\includegraphics[width=0.495\linewidth]{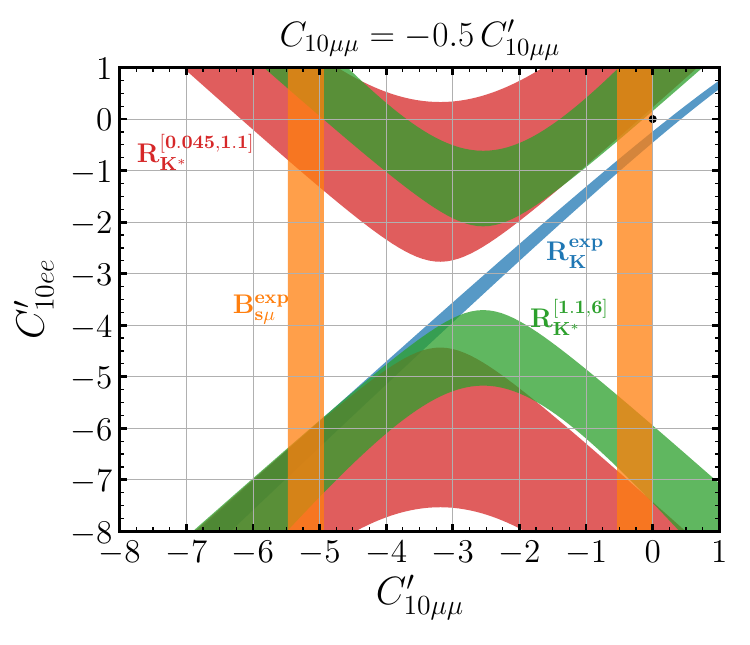}
\includegraphics[width=0.495\linewidth]{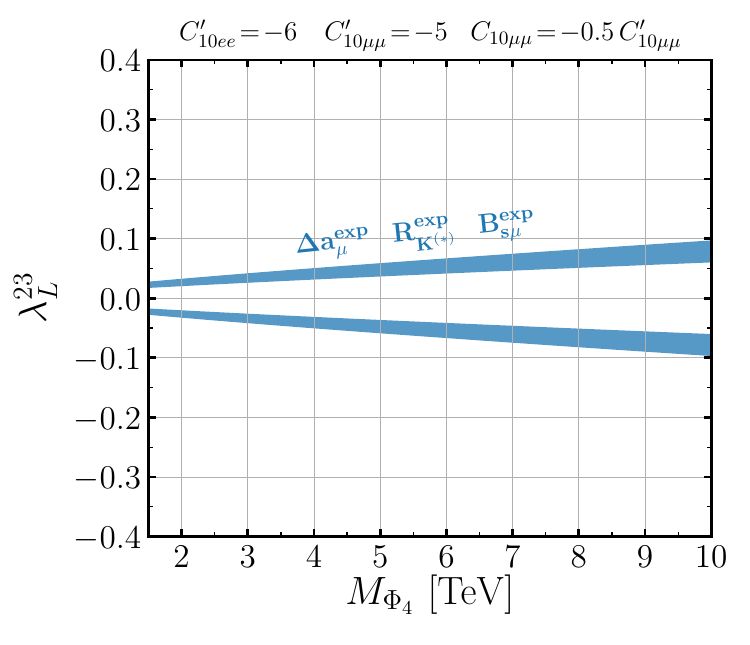}
\caption{\textit{Left panel}: Same as Fig.~\ref{fig:ParameterSpacePhi3}; we also include a contribution from $C_{10\mu\mu} = -0.5 \, C'_{10\mu\mu}$. \textit{Right panel}: The region shaded in blue is in agreement with the combined result from the Muon $g−2$ experiment at Fermilab and E821 at BNL within $1\sigma$. We have also fixed the Wilson coefficients that reproduce the experimental measurements of $\text{Br}(B_s \to \mu^+ \mu^-)$ and $\mathcal{R}_K$ within $1\sigma$, and $\mathcal{R}_{K^*}$ within $2\sigma$.  }
\label{correlation}
\end{figure}

In the left panel in Fig.~\ref{correlation} we show the parameter space of the Wilson coefficients $C^{'}_{10ee}$, $C^{'}_{10\mu\mu}$, and $C_{10\mu\mu}$, 
assuming that $\Phi_{4}$ couples mainly to muons (i.e. adopting the texture in Eq.~\eqref{eq:textureV4}). As can be seen in this plot, the flavor observables can be reproduced with $C^{'}_{10ee}\!\approx\!-6$, $C^{'}_{10\mu\mu}\!\approx\!-5$, and for the contribution from $\Phi_4$ we take $C_{10\mu\mu}=-0.5 \, C^{'}_{10\mu\mu}$.

The entries in the $V_6$ matrix can be written in terms of the coefficient $C_{10\mu\mu}$ as follows
\begin{equation}
{V}_6^{32} ({V}_6^{22})^* = -4 \, M^2_{\phi_4^{2/3}} C_{10\mu\mu} \left( \frac{G_F V_{tb} V_{ts}^* \ \alpha}{{\pi \sqrt{2}}} \right),
\end{equation}
where we have assumed $K_2 V_{\rm CKM} K_1 \sim \mathbf{1}$, for simplicity. Notice that one of the couplings, $\lambda_R^{23}$, needed to predict $(g-2)_\mu$ 
is $\lambda_R^{23}=-V_6^{32}$. Then, assuming ${V}_6^{32} = -{V}_6^{22}$, the coupling $\lambda_R^{23}$ can be written as a function of $M_{\phi_4^{2/3}}$ using the above equation. Neglecting the mass splitting between the fields in $\Phi_4$, we have $M_{\phi_4^{2/3}}=M_{\phi_4^{5/3}}=M_{\Phi_4}$. Therefore, the predictions for $g-2$ will depend only on 
two parameters, $\lambda_L^{23}$ and $M_{\Phi_4}$. In the right panel in Fig.~\ref{correlation} we show the predictions for the Muon $g-2$ after fixing the Wilson coefficients to the values that reproduce the flavor anomalies. The blue band is in agreement with the measured value of $(g-2)_\mu$ within $1\sigma$. Consequently, there are consistent scenarios that provide a simultaneous explanation of the flavor and muon $g-2$ anomalies.

\begin{figure}[t]
\centering
\includegraphics[width=0.495\linewidth]{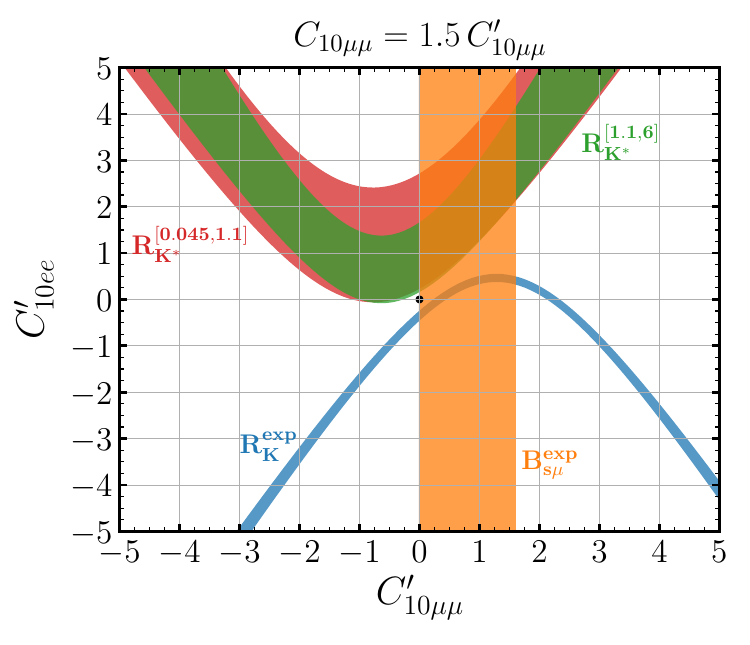}
\includegraphics[width=0.495\linewidth]{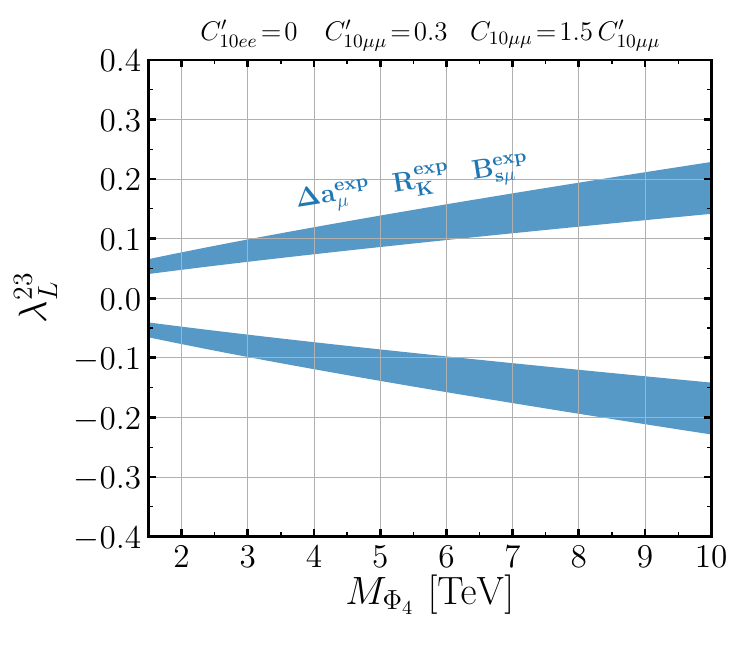}
\caption{\textit{Left panel}: Same as Fig.~\ref{fig:ParameterSpacePhi3}; we also include a contribution from $C_{10\mu\mu} = 1.5 \, C'_{10\mu\mu}$. \textit{Right panel}: The region shaded in blue is in agreement with the combined result from the Muon $g−2$ experiment at Fermilab and E821 at BNL within $1\sigma$. We have also fixed the Wilson coefficients that reproduce the experimental measurements of $\text{Br}(B_s \to \mu^+ \mu^-)$ and $\mathcal{R}_K$ within $1\sigma$. The predicted values for $\mathcal{R}_{K^*}$ are higher than the current central values as discussed in the text. } 
\label{correlation2}
\end{figure}

There also exists a solution with smaller values for the Wilson coefficients even though the measured values for $\mathcal{R}_{K^*}$ cannot be reproduced. First of all, for $\phi_4^{5/3}$ to be able to explain $(g-2)_\mu$ at the multi TeV scale we need the coefficient $C_{10\mu\mu}$ to be order one, so we set $C_{10\mu\mu}=1.5 \, C'_{10\mu\mu}$. From the left panel in Fig.~\ref{correlation2} we see that for $C'_{10\mu\mu}=0.3$ and $C'_{10e e}=0$ we can reproduce the measured values for $\mathcal{R}_K$ and ${\rm Br}(B_s \to \mu^+ \mu^-)$; moreover, in this scenario the new physics is only coupled to muons. However, the predictions for  $\mathcal{R}_{K^*}$ turn out to be larger than the current observed value. In the right panel of Fig.~\ref{correlation2} we assume that $V_6^{32} = -V_6^{22}$ and show the region in the $\lambda_L$ vs $M_{\Phi_4}$ plane that explains the $g-2$ of the muon within $1\sigma$.

Focusing on the region that reproduces $\text{Br}(B_s \to \mu^+ \mu^-)$ and $\mathcal{R}_K$ within $1\sigma$, in the left panel of Fig.~\ref{correlation2}, we obtain the following predictions for $\mathcal{R}_{K^*}$: $ \mathcal{R}_{K^*}^{[0.045,1.1]}  = [0.98 \, -  \, 1.34]$, $\mathcal{R}_{K^*}^{[1.1,6]}  =  [1.08 \, - \, 1.85]$, both are much higher than the current observed values given in Eq.~\eqref{eq:RKstarEXP}. The values of the Wilson coefficients on the right panel in Fig.~\ref{correlation2} correspond to the lower end in these ranges.\clearpage
\section{SUMMARY}
\label{sec:summary}
We have discussed the simplest quark-lepton unification theory that can be realized at the TeV scale~\cite{Perez:2013osa} and can be seen as a low energy limit of the Pati-Salam theory. This theory is based on the $\SU(4)_C \otimes \SU(2)_L \otimes \U(1)_R$ gauge group and, in order to have a consistent theory for fermion masses at the low scale, neutrino masses are generated through the inverse seesaw mechanism. This theory predicts the existence of a vector leptoquark, $X_\mu \sim (\mathbf{3}, \mathbf{1},2/3)_{\rm SM}$, and two scalar leptoquarks, $\Phi_3 \sim (\mathbf{\bar{3}}, \mathbf{2}, -1/6)_{\rm SM}$ and $\Phi_4\sim (\mathbf{3}, \mathbf{2}, 7/6)_{\rm SM}$, that can provide a relevant contribution to meson decays.

We have studied the possibility to explain the experimental values for the clean observables involving $b \to s$ transitions, i.e. $\mathcal{R}_K$, $\mathcal{R}_{K^*}$ and $\text{Br}(B_s \to \mu^+ \mu^-)$, in two main scenarios. In the first scenario the scalar leptoquark $\Phi_3$ gives the main contributions to explain the measured values of the relevant meson decays through couplings to both electrons and muons. In the second scenario the scalar leptoquark $\Phi_4$ plays the main role to explain the values for the neutral flavor anomalies; in this scenario the New Physics is coupled mostly to electrons as it is required by the experimental bound from $\mu \to e \gamma$. Furthermore, we showed that $\Phi_4$ can be used to explain the $g-2$ of the muon while being consistent with other experimental bounds. 

We found scenarios where we can address simultaneously the flavor and the $(g-2)_\mu$ anomalies, in which both leptoquarks $\Phi_3$ and $\Phi_4$ play a role. In these scenarios, the recent experimental results for $\mathcal{R}_K$ and $\mathcal{R}_{K^*}$ are explained by contributions from $\Phi_3$ and $\Phi_4$, with Wilson coefficients of the same order as in the SM, while the measured value of $(g-2)_\mu$ can be addressed by coupling $\Phi_4$ mostly to muons, so that the aforementioned anomalies can all be explained in consistency with constraints from lepton flavor violation.  

We hope that, in the near future, more experimental data and an improvement on the theoretical predictions will determine whether these anomalies represent final evidence for New Physics, and whether the minimal theory for quark-lepton unification can be behind them by contrasting alternative predictions with experimental results.

\vspace{0.25cm}
{\textit{Acknowledgments:}}
{\small We thank the referee for important feedback on an earlier version of this work. The work of P.F.P. has been supported by the U.S. Department of Energy,
Office of Science, Office of High Energy Physics, under Award Number 
DE-SC0020443.
The work of C.M. is supported by the U.S. Department of Energy, Office of Science, Office of High Energy Physics, under Award Number DE-SC0011632 and by the Walter Burke Institute for Theoretical Physics.}

\appendix

\section{Leptoquark Interactions}
\label{interactions}
In our convention the mass matrices are diagonalized as
\begin{eqnarray}
&& U^T M_U U_C = M_U^{\rm diag}, \\
&& D^T M_D D_C = M_D^{\rm diag}, \\
&& E^T M_E E_C = M_E^{\rm diag}.
\end{eqnarray}

The following matrices enter in the leptoquark interactions below:

\vspace{0.3cm}

\begin{center}
$V_1=N_C^\dagger U_C$, $V_2=E_C^\dagger D_C$, $V_3=U^T Y_2 N_C$, $V_4=N^T Y_4 D_C$, $V_5=N^T Y_2 U_C$, and $V_6=U^T Y_4 E_C$.
\\
\vspace{0.3cm}
$V_{DE} =D^\dagger E$, \,\,\, $U^\dagger D = K_1 V_{\rm CKM} K_2$  \,\,\, and \,\,\, $E^\dagger N = K_3 V_{\rm PMNS}$.
\end{center}
$K_1$ and $K_3$ are diagonal matrices containing three phases, while $K_2$ has two phases.

\begin{itemize}

\item Vector Leptoquark $X_\mu \sim (\mathbf{3}, \mathbf{1}, 2/3)_{\rm SM}$: 

\begin{eqnarray}
&& \frac{g_4}{\sqrt{2}} \bar{d}_L \ V_{DE} \ \gamma^\mu e_L X_\mu, \\
&&  \frac{g_4}{\sqrt{2}} \bar{u}_L \ ( K_1 V_{\rm CKM} K_2 V_{DE} K_3 V_{\rm PMNS} ) \ \gamma^\mu \nu_L X_\mu, \\
&&  \frac{g_4}{\sqrt{2}} \overline{(\nu^c)}_L \ V_1 \ \gamma^\mu (u^c)_L X_\mu, \\
&&  \frac{g_4}{\sqrt{2}} \overline{(e^c)}_L \ V_2 \ \gamma^\mu (d^c)_L X_\mu. 
\end{eqnarray}

\item Scalar Leptoquark $\Phi_3 \sim (\mathbf{\bar{3}}, \mathbf{2}, -1/6)_{\rm SM}$:

\begin{eqnarray}
&& u_L^T C \ V_3 \ (\nu^c)_L \ \phi_3^{-2/3}, \\
&& -d_L^T C \ K_2 V_{\rm CKM}^T K_1 V_3 \ (\nu^c)_L \ \phi_3^{1/3}, \\
&& \nu_L^T C \ V_4 \ (d^c)_L \  (\phi_3^{1/3})^*,\\
&& e_L^T C \ K_3^* V_{\rm PMNS}^* V_4 \ (d^c)_L \  (\phi_3^{-2/3})^*.
\end{eqnarray}\\ \\

\item Scalar Leptoquark $\Phi_4 \sim (\mathbf{3}, \mathbf{2}, 7/6)_{\rm SM}$:

\begin{eqnarray}
&& \nu_L^T C V_5 (u^c)_L \phi_4^{2/3}, \\
&& - e_L^T C \ K_3^* V_{\rm PMNS}^* V_5 \ (u^c)_L \  \phi_4^{5/3}, \\
&& u_L^T C \ V_6 \ (e^c)_L \  (\phi_4^{5/3})^*,\\
&& d_L^T C \ K_2 V_{\rm CKM}^T K_1 V_6 (e^c)_L (\phi_4^{2/3})^*.
\end{eqnarray}

\end{itemize}

Notice that when $Y_2 \to 0$ the matrices $V_3 \to 0$ and $V_5 \to 0$.

\section{Bounds from $\mu \to e \gamma$}
\label{appMEG}
\begin{figure}[t]
\centering
\includegraphics[width=0.5\linewidth]{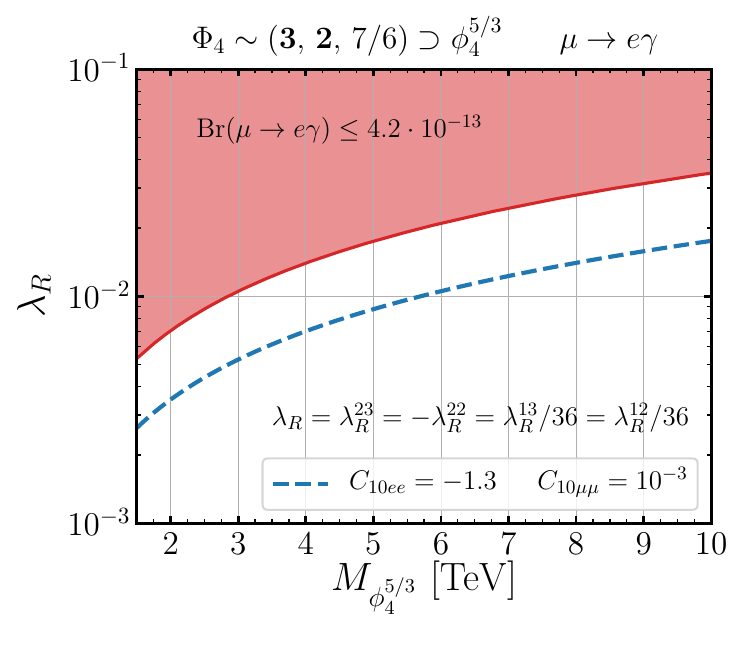}
\caption{Parameter space in the $\lambda_R$ vs $M_{\phi_4^{5/3}}$ plane, the region shaded in red is excluded by the experimental bound ${\rm Br}(\mu \to e \gamma)\leq 4.2 \times 10^{-13}$~\cite{MEG:2016leq}. The other couplings have been fixed to $\lambda_R = \lambda_R^{23} = -\lambda_R^{22} = \lambda_R^{13}/36 = \lambda_R^{12}/36$ so that we obtain $C_{10\mu\mu}=10^{-3}$ and $C_{10ee}=-1.3$ which are able to explain the flavor anomalies.
}
\label{fig:MEG}
\end{figure}
In this appendix we discuss the implications from the experimental bounds on $\mu \to e \gamma$ on the three scenarios studied in this work:

\begin{itemize}

\item ${\bf \boldmath \Phi_3}$ {\bf scenario}: In this case only the field $\phi_3^{-2/3}$ contributes to $\mu \to e \gamma$ but its contribution is chiral suppressed. Furthermore, the contribution from $\phi_3^{-2/3}$ has a near cancellation in the loop functions due to the electric charge of $-2/3$; this is because in the limit $x \to 0$ the loop functions approach $F_1(x) \to 1/3$ and $F_2(x)\to1/6$, and hence, the combination $Q_d F_1(x) - Q_{\rm LQ} F_2(x) \approx {\cal O}(x)$, i.e. it almost vanishes.

\item ${\bf \boldmath \Phi_4}$ {\bf scenario}: In this case two fields, $\phi_4^{2/3}$ and $\phi_4^{5/3}$, contribute to $\mu \to e \gamma$. The contribution from  $\phi_4^{2/3}$ is suppressed as in the case 
of $\phi_3^{-2/3}$ discussed above. However, the component $\phi_4^{5/3}$ can still give a contribution to this observable larger than the current experimental bound.
The decay width is given by~\cite{Lavoura:2003xp,Benbrik:2008si},
\beq
\Gamma(\mu \to e \gamma) \simeq \frac{\alpha}{4} \frac{m_\mu^5}{M_{\phi_4^{5/3}}^4}  \sum_j   \left| \frac{3}{32\pi^2}    \,  \lambda_R^{2j} \lambda_R^{1j*} \left[ Q_q F_1(x_j) + Q_{\rm LQ} F_2(x_j) \right]\right|^2  \, ,
\eeq
with $x_j=\left({m_{u_j}}/{M_{\phi_4^{5/3}}}\right)^2$, $\lambda_R=V_6^T$ and assuming $V_\text{CKM} \sim \mathbf{1}$ we have that $\tilde{V}_6 \simeq V_6$.
In Fig.~\ref{fig:MEG} we show the parameter space in the $\lambda_R$ vs $M_{\phi_4^{5/3}}$ plane. The region shaded in red is excluded by the experimental bound ${\rm Br}(\mu \to e \gamma)\leq 4.2 \times 10^{-13}$~\cite{MEG:2016leq}. This is the motivation behind the texture chosen in Eq.~\eqref{eq:Y4texturePhi4} with couplings mostly to electrons. For the plot we have chosen the benchmark values of $\lambda_R = \lambda_R^{23} = -\lambda_R^{22} = \lambda_R^{13}/36 = \lambda_R^{12}/36$, so that we obtain $C_{10\mu\mu}=10^{-3}$ and $C_{10ee}=-1.3$, which are able to explain the flavor anomalies. We are also taking $M_{\Phi_4}=M_{\phi_4^{5/3}}=M_{\phi_4^{2/3}}$ since the mass splitting cannot be large.

\item ${\bf \boldmath \Phi_3 \,\, \textbf{\&} \,\, \Phi_4}$ {\bf scenario}: In Section~\ref{all} we discussed the scenario where both fields, $\Phi_3$ and $\Phi_4$, contribute to the flavor anomalies and the connection between the predictions for $\mathcal{R}_K$ and $(g-2)_\mu$. In this case, the bound from $\mu \to e \gamma$ can be satisfied because the coupling of $\Phi_4$ to electrons is suppressed and the $\Phi_3$ contribution to such processes is also suppressed, and therefore, this bound can be neglected as in the first case discussed above.

\end{itemize}

\FloatBarrier

\bibliography{main}

\end{document}